\documentclass[aps,prd,preprint,groupedaddress,amssymb,amsmath,nofootinbib]{revtex4}

\usepackage{color}
\usepackage{graphicx}
\usepackage{epstopdf}
\usepackage{hyperref} 

\usepackage{slashed} 

\begin{document}

\preprint{}

\title{Z' near the Z-pole}

\author{Radovan Derm\' \i\v sek}
\email[]{dermisek@indiana.edu}
\affiliation{Physics Department, Indiana University, Bloomington, IN 47405, USA}

\author{Sung-Gi Kim}
\email[]{kimsg@indiana.edu}
\affiliation{Physics Department, Indiana University, Bloomington, IN 47405, USA}

\author{Aditi Raval}
\email[]{adiraval@indiana.edu}
\affiliation{Physics Department, Indiana University, Bloomington, IN 47405, USA}


\date{December 31, 2011}

\begin{abstract}

We present a fit to precision electroweak data in the standard model extended by an additional vector boson, $Z'$, with suppressed couplings to the electron compared to the $Z$ boson,  with couplings to the b-quark, and with mass close to the mass of the $Z$ boson. This scenario provides an excellent fit to forward-backward asymmetry of the b-quark measured on the $Z$-pole and $\pm 2$ GeV off the $Z$-pole, and to lepton asymmetry, $A_e$, obtained from the measurement of left-right asymmetry for hadronic final states, and thus
it removes the tension in the determination of the weak mixing angle from these two measurements.
It also leads to a significant improvement in the total hadronic cross section on the $Z$-pole and $R_b$ measured at energies above the $Z$-pole. 
We explore in detail properties of the $Z'$ needed to explain the data and
present a model for $Z'$ with required couplings. 
The model preserves standard model Yukawa couplings, it is anomaly free and can be embedded into grand unified theories. It allows a choice of parameters that does not generate any flavor violating couplings of the $Z'$ to standard model fermions. Out of standard model couplings, it  only  negligibly modifies the left-handed bottom quark coupling to the $Z$ boson and the 3rd column of the CKM matrix. Modifications of standard model couplings in the charged lepton sector are also negligible. It predicts an additional down type quark, $D$, with mass in a few hundred GeV range, 
and an extra lepton doublet, $L$, possibly much heavier than the $D$ quark.
We discuss signatures of the $Z'$ at the Large Hadron Collider and calculate the $Z'b$ production cross section
which is the dominant production mechanism for the $Z'$. 

\end{abstract}

\pacs{}
\keywords{}

\maketitle






\section{Introduction}

Among the largest deviations from  predictions of the standard model (SM)
 is the discrepancy in determination of the weak mixing angle from the LEP measurement of the forward-backward asymmetry of the b-quark,  $A^b_{FB}$,  
and from  the SLD measurement of left-right asymmetry for hadronic final states, $A_e{(\rm LR-had.)}$.
These two measurements, showing the  largest deviations from SM predictions among Z-pole observables,  create a very puzzling situation~\cite{Chanowitz:2002cd},\cite{Nakamura:2010zzi}. 
Varying  SM input parameters, especially the Higgs boson mass, one can fit the experimental value for one of them only at the expense of increasing the discrepancy in the other one. While $A^b_{FB}$ prefers a heavy Higgs boson, $m_h \simeq 400$ GeV,  $A_e{(\rm LR-had.)}$ prefers  $m_h \simeq 40$ GeV. Since other observables also prefer a lighter Higgs the focus has been on possible new physics effects that modify $A^b_{FB}$. However, if the pull for a large Higgs mass from $A^b_{FB}$ is removed, the global fit preference 
is in tension with LEP exclusion limit,  $m_h > 114$ GeV~\cite{Barate:2003sz}. 

In a previous study~\cite{Dermisek:2011xu} we showed that a $Z'$ with mass close to the mass of the $Z$ boson, with suppressed couplings to the electron compared to the Z-boson,  and with couplings to the b-quark,  
provides an excellent fit to measurements of $A^b_{FB}$ on and near the $Z$-pole,  and simultaneously to $A_e{(\rm LR-had.)}$.
It also leads to a significant improvement in the total hadronic cross section on the Z-pole and $R_b$ measured at energies above the $Z$-pole. In addition, with a proper mass, the $Z'$ can explain the $2.3\sigma$ excess of $Zb\bar b$ events  at LEP in the $90-105$ GeV region of the $b\bar b$ invariant mass, thus expanding the family of possible explanations of the excess that include a Higgs boson with reduced coupling to the $Z$ boson~\cite{Sopczak:2001tk,Carena:2000ks,Drees:2005jg} or
a SM-like Higgs boson with reduced branching fraction to $b \bar b$~\cite{Dermisek:2005ar,Dermisek:2005gg,Dermisek:2007yt}.

In this paper, we explore in detail properties of the $Z'$ needed to explain the data and
present a model for $Z'$ with required couplings. We discuss  signatures of the model at the Large Hadron Collider. We calculate the $Z'b$ production cross section which is the dominant production mechanism for the $Z'$ and discuss signatures of extra vector-like quarks that are predicted by the model.

We consider a new vector boson, $Z'$, associated with a new gauge symmetry $U(1)'$, with couplings 
 to the electron and the b-quark:
\begin{equation}
{\cal L} \supset    Z'_\mu  \bar e \gamma^\mu (g'^e_L P_L  + g'^e_R P_R ) e +
  Z'_\mu \bar b \gamma^\mu (g'^b_L P_L  + g'^b_R P_R ) b.
 \vspace{-0.1cm}
\label{eq:lagrangian}
\end{equation}
In the numerical analysis we do not make
 any assumptions about the origin of the $Z'$ and treat all four couplings and the mass of the $Z'$  as free parameters. 
Couplings to other SM fermions and the mixing with the $Z$ boson are assumed to be negligible and are set to zero for simplicity.  
Once we determine typical  sizes of $g'^{e,b}_{L,R}$ couplings required we construct a model that generates them through mixing of standard model fermions with extra vector-like fermions charged under the $U(1)'$. This method of effectively generating $Z'$ couplings was recently discussed in Ref.~\cite{Fox:2011qd}.
Although this is not the only possible model, it is a simple one that preserves standard model Yukawa couplings, it is anomaly free and can be embedded into grand unified theories.

Previous explanations of the deviation in  $A^b_{FB}$ focused on modifying $g_R^b$. 
Achieving this and simultaneously not upsetting quite precise agreement in $R_b$ turned out to be very challenging for a new physics that enters through loop corrections~\cite{Haber:1999zh}. This motivated  tree level modification of the $g_R^b$ either through mixing of b-quark with extra fermions~\cite{Choudhury:2001hs} or through Z-Z' mixing~\cite{He:2002ha,delAguila:2010mx}. 
However the $A^b_{FB}$ is only a part of the puzzle and  any new physics that reduces to modification of bottom quark couplings cannot affect the $A_e{(\rm LR-had.)}$.

In Ref.~\cite{Dermisek:2011xu} we suggested to modify the $b \bar b$ production cross section directly, $e^+e^- \to Z'^* \to b \bar b$, rather than modifying the $Z$-couplings. 
This idea comes from a simple observation that increasing $\sigma(e_L \bar e_L \to b_R \bar b_R)$ can decrease $A^b_{FB}$ and simultaneously increase $A_e{(\rm LR-had.)}$ by  a relative factor that is needed to bring them close to observed values while still improving on $R_b$ (for more details see Ref.~\cite{Dermisek:2011xu}).  This can be only achieved by a $Z'$ near the $Z$-pole.
To generate a sizable contribution to $A^b_{FB}$ while contributing to $R_b$ only negligibly on the $Z$-pole, and not significantly affect predictions for $A^b_{FB}$ and $R_b$ above the $Z$-pole (that roughly agree with measurements), the increase in $\sigma(e_L \bar e_L \to b_R \bar b_R)$ must be due to  the s-channel exchange of a new vector particle with mass close to the mass of the $Z$ boson. A
 scalar particle near the $Z$-pole can modify $A^b_{FB}$  only comparably to its modification of $R_b$.\footnote{This was considered in Ref.~\cite{Erler:1996ww} motivated by previous discrepancies in $Z$-pole observables, namely a large deviation in $R_b$ (which currently agrees with the SM prediction).} 
A heavy  particle, or a particle contributing in t-channel, can modify $Z$-pole observables only negligibly if it should not dramatically alter predictions above the $Z$-pole.
Thus a $Z'$ near the $Z$-pole with small couplings to the electron (in order to satisfy limits from searches for $Z'$) and sizable couplings to the bottom quark is the only candidate. 

There is extensive literature concerning models for $Z'$ and their phenomenological implications~\cite{Langacker:2008yv}.
A $Z'$ was frequently used to explain previous discrepancies in precision electroweak data, see {\it e.g.} a heavy $Z'$~\cite{Erler:1999nx} or almost degenerate $Z$ and $Z'$~\cite{Caravaglios:1994jz} scenarios. Related constraints on a $Z'$ near the $Z$-pole were discussed in Refs.~\cite{Babu:1996vt,Babu:1997st}.

This paper is organized as follows. In Sec. II we outline the numerical analysis. The results of the best fit to precision electroweak data and ranges of $Z'$ mass and couplings needed to fit the data are presented in Sec. III. A possible model leading to required couplings is discussed in Sec. IV. The current constraints and LHC predictions are summarized in Sec. V,  and we conclude in Sec.~VI.

\section{Numerical analysis}

We construct a $\chi^2$ function of relevant quantities related to the bottom quark and electron measured at and near the $Z$-pole which are summarized in Table~\ref{tab:fit}. Their precise definition can be found in the EWWG review~\cite{:2005ema} from which we also take the corresponding experimental values.
Instead of the pole  forward-backward asymmetry of  the b-quark, $A_{FB}^{0,b}$, we include three measurements of the asymmetry, at the peak and $\pm 2$ GeV from the peak. These are more relevant because the presence of a $Z'$ near the $Z$-pole changes the energy dependence of the asymmetry. In addition,  about 25\% of the deviation in the pole asymmetry comes from the measurement at $+2$ GeV from the peak. Corresponding LEP averages for $R_b$ at  $\pm 2$ GeV from the peak do not exist. These are available only from DELPHI~\cite{Abreu:1998xf} and although they are included in the $Z$-pole LEP average,  $R_b^0$, we include them in addition in order to constrain the energy dependence.
 We further  include pole values of the total hadronic cross section, $\sigma^0_{\rm had}$, the ratio of the hadronic and electron decay widths, $R_e^0$, forward-backward asymmetry of  the electron, $A_{FB}^{0,e}$, measured at LEP; and  SLD values of asymmetry parameters of the b-quark,   $A_b$, obtained from the measurement of left-right forward-backward asymmetry, and the electron, obtained from the measurement of left-right asymmetry for hadronic final states, $A_e{(\rm LR-had.)}$, and leptonic final states, $A_e{(\rm LR-lept.)}$.  
 
 Although we do not modify production cross sections for the charm quark and other charged leptons we nevertheless include related 
 observables  in the $\chi^2$ because of correlations with observables related to the bottom quark and the electron. The correlations are included for the following two sets of observables.
The first set  consists of 9 pseudo-observables: $m_Z$, $\Gamma_Z$, $\sigma_{\rm had}^0$, $R_e^0$, $R_\mu^0$, $R_\tau^0$, $A_{FB}^{0,e}$, $A_{FB}^{0,\mu}$, $A_{FB}^{0,\tau}$.
The second set represents 18 heavy-flavor observables:
$R_b^0$,
$R_c^0$, 
$A_{FB}^b (-2)$,
$A_{FB}^c (-2)$,
$A_{FB}^b ({\rm pk})$,
$A_{FB}^c ({\rm pk})$, 
$A_{FB}^b (+2)$, 
$A_{FB}^c (+2)$, 
 $A_b$,
 $A_c$, 
and 8 additional  b- and c-tag related observables that we fix to the fitted values.
Precise definition of these observables can be found in the EWWG review~\cite{:2005ema} from which we also take the corresponding experimental values and correlations.

While b-quark quantities were measured at three energies near the $Z$-pole, the total hadronic cross section was measured also at $\pm 1,3$ GeV (from data collected only during 1990-1991) by four LEP collaborations \cite{Barate:1999ce, Abreu:2000mh, Acciarri:2000ai, Abbiendi:2000hu}.
Because there are no combined results, we take ALEPH results to estimate the relative errors for each measured $\sigma_{\rm had}(\sqrt{s})$ as 1.8\%, 0.4\%, 1.1\%, 1.1\%, 0.3\%, 1.3\% for $\sqrt{s} = -3, -2, -1, 1, 2, 3$ GeV from the peak respectively.%
\footnote{ALEPH collaboration quotes both statistical and systematic errors and the combined errors  are comparable to just statistical errors quoted by other LEP collaborations.}
We then require the total hadronic cross section including Z' to deviate from the SM cross section no more than twice the estimated experimental error at a  given $\sqrt{s}$.

We calculate theoretical predictions using ZFITTER 6.43~\cite{Bardin:1999yd,Arbuzov:2005ma} and  ZEFIT 6.10~\cite{Leike:1991if} which we modified for a $Z'$ with free couplings to the b-quark and the electron.  
In the case of the standard model we precisely reproduce the result in the EWWG review~\cite{:2005ema} or the PDG review~\cite{Nakamura:2010zzi} for sets of SM input parameters used in those fits. 
In our fit we use the SM input parameters summarized in Table 8.1 of the EWWG review~\cite{:2005ema}, namely: $m_Z = 91.1875$ GeV, $\Delta \alpha^{(5)}(m_Z^2) = 0.02758$, $\alpha_S(m_Z^2) = 0.118$; however, we update the  top quark mass  to the Tevatron average, $m_t = 173.3$ GeV~\cite{:1900yx}, and fix the Higgs mass to $m_H = 117$ GeV. 
As a result of the different set of SM input parameters, our SM predictions, given in Table~\ref{tab:fit}, are slightly different from~\cite{:2005ema,Nakamura:2010zzi}. The effects of varying input parameters on electroweak observables can be found in~\cite{:2005ema}. The differences resulting from a given choice of SM input parameters are not essential for comparison of the SM and SM+$Z'$.
We minimize the $\chi^2$ function of 5 parameters, $m_{Z'}$, $g'^e_L$, $g'^e_R$, $g'^b_L$, and $g'^b_R$, with MINUIT~\cite{minuit}. In principle, the width,  $\Gamma_{Z'}$, could be treated  as a free parameter because $Z'$ can have additional couplings 
that do not affect precision electroweak data. For simplicity, we do not consider this possibility.

\section{Results}

\begin{figure}[t] 
   \centering
   \includegraphics[width=2.8in]{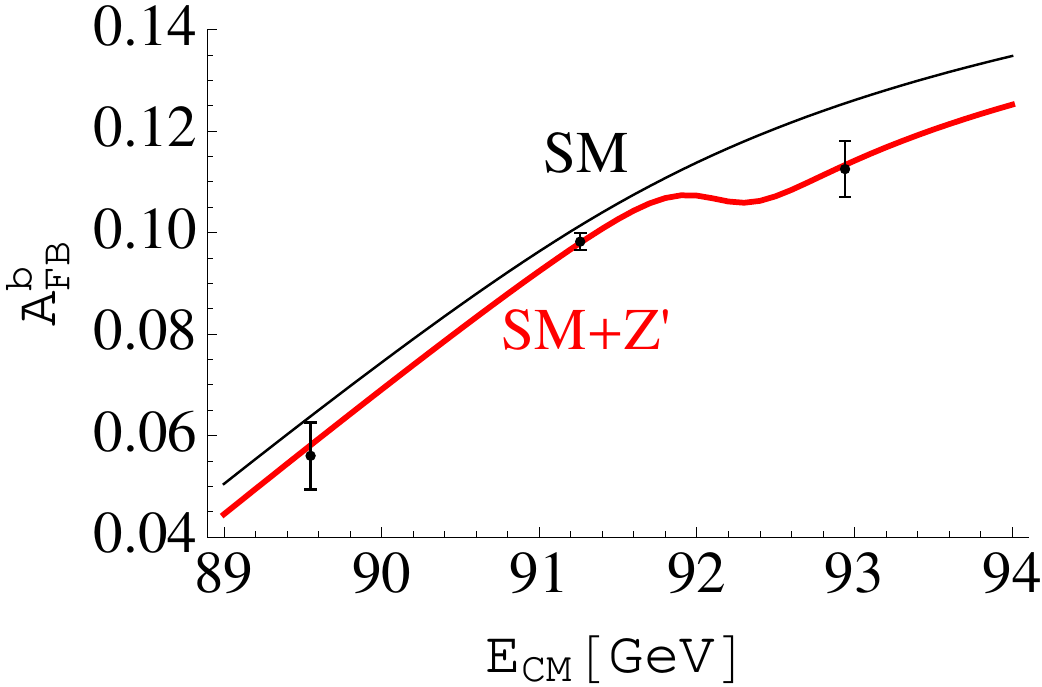}      
    \includegraphics[width=2.8in]{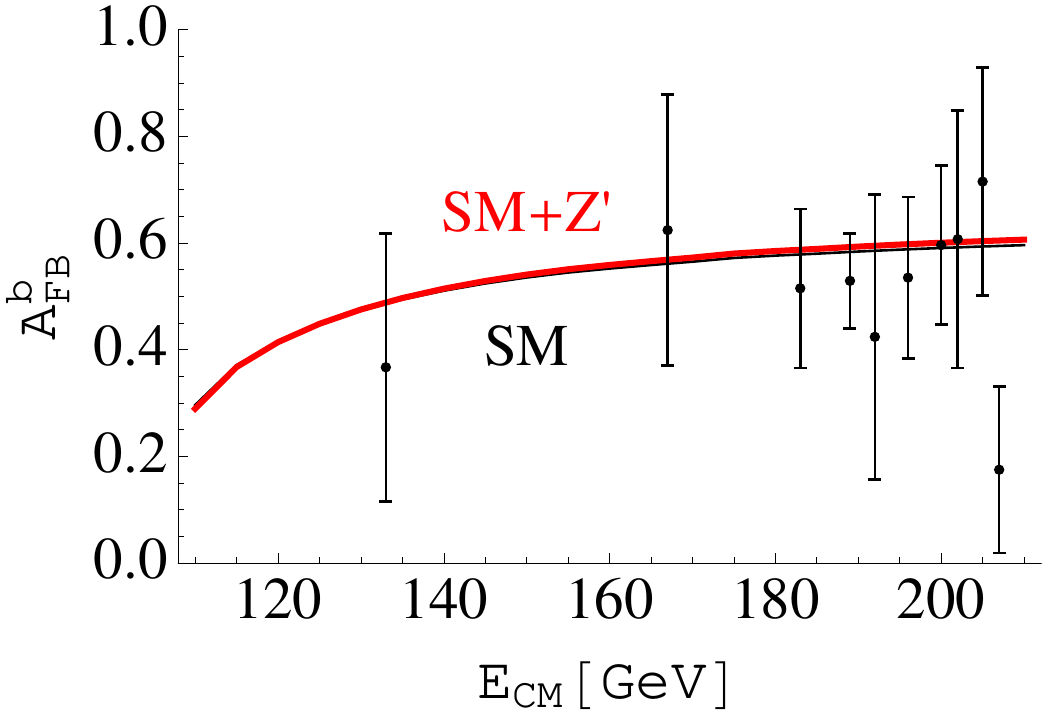}\\
   \includegraphics[width=2.8in]{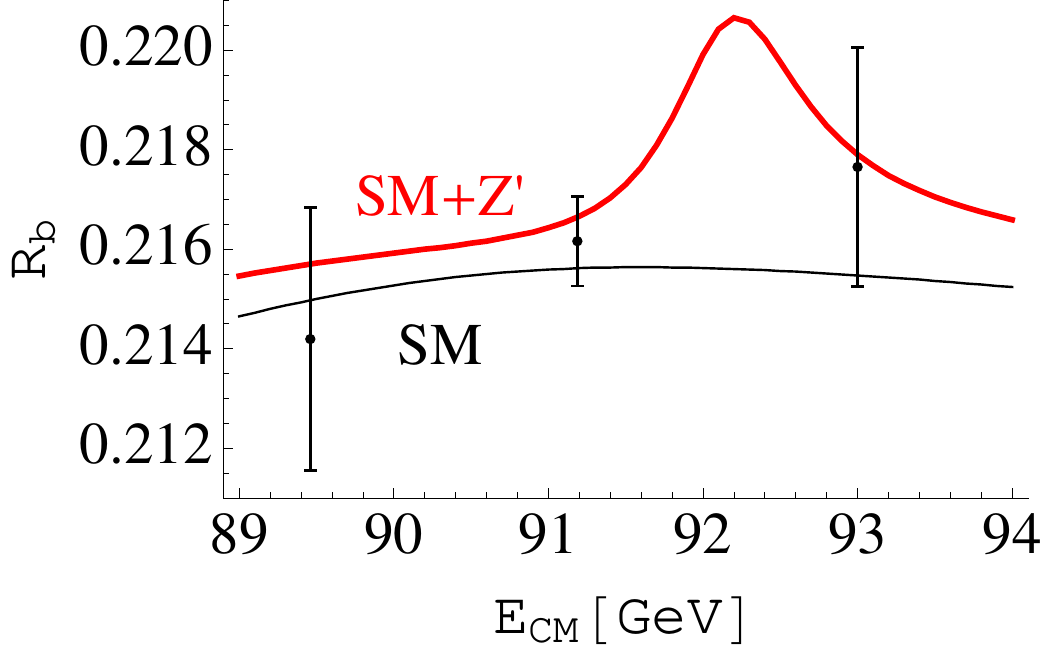}  
       \includegraphics[width=2.8in]{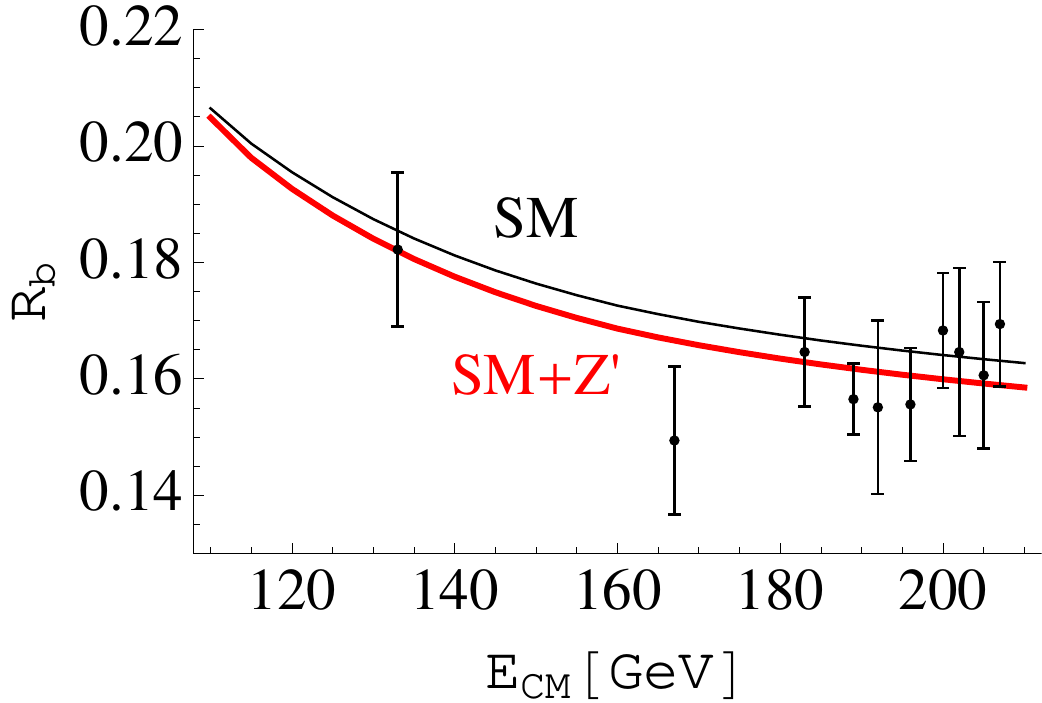} 
   \caption{Experimental values of $A_{FB}^b$ (top) and $R_b$ (bottom) and predictions of  the SM (thin lines) and the Z' model (thick lines) for  input parameters corresponding to the best fit I specified in the caption of Table~\ref{tab:fit} as functions of center of mass energy near and above the Z-pole.}
   \label{fig:NP}
\end{figure}

The best fits to precision data included in the $\chi^2$ are summarized  in Table~\ref{tab:fit} and  parameters for which the best fits are obtained are given in the caption. The fit I corresponds to all four couplings in Eq.~(\ref{eq:lagrangian}) allowed to vary, while the fit II assumes that only two couplings, $g'^e_L$ and $g'^b_R$, are free parameters, and $g'^e_R = g'^b_L = 0$. Best fits are also compared with predictions of the standard model.
Clearly, addition of a $Z'$ provides an excellent fit to selected precision electroweak data with $\chi^2 = 6.8$ 
with 5 additional parameters compared to the standard model that has $\chi^2 = 25.$\footnote{The difference in $\chi^2$ compared to Ref.~\cite{Dermisek:2011xu} is the result of a more complete $\chi^2$ function that includes more  observables and correlations between them. The best fit values of $Z'$ parameters and main features of results are not affected by these modifications.} The most significant improvement comes from the three measurements of $A_{FB}^b$ which can be fit basically at central values in the $Z'$ model, without spoiling the agreement in $R_b$. The energy dependence of both quantities near the $Z$-pole  for both the SM and the $Z'$ model together with data points is plotted in Fig.~\ref{fig:NP}.
The $A_e{(\rm LR-had.)}$ and $\sigma_{\rm had}^0$ are also fit close to their central values. The fit II illustrates that most of the improvement originates from two couplings $g'^e_L$ and $g'^b_R$ which was already discussed in Ref.~\cite{Dermisek:2011xu}. Allowing all four couplings further improves the fit and also enlarges the ranges of couplings for which a good fit is achieved.

\begin{table}[t]
    \caption{The best fits to relevant precision electroweak observables in the SM with a $Z'$.
    The best fit I  assumes all 4 couplings are free parameters  and it is achieved for: $m_{Z'} = 92.2$ GeV, 
    $g'^e_L = 0.0065$, $g'^e_R = 0.0077$, $g'^b_L = 0.044$,  and $g'^b_R = -0.51$; ($\Gamma_{Z'} = 1.0$ GeV).
    The best fit II  assumes only two couplings are free parameters ($g'^e_R = g'^b_L = 0$) and it is achieved for: $m_{Z'} = 92.2$ GeV, 
    $g'^e_L = 0.0048$,  and $g'^b_R = -0.52$; ($\Gamma_{Z'} = 1.0$ GeV).
    The standard model input parameters are fixed to $m_t = 173.3$ GeV, $m_h = 117$ GeV, and other parameters as listed in Table 8.1 of the  EWWG review~\cite{:2005ema}. For comparison, we also include predictions of the standard model with $\chi^2$ contributions.
    }

\begin{tabular}{  l  c  c c c c c c c c c}
 \hline
  \hline
Quantity    & Exp. value & {\color{white}{{--}}} & SM& $\chi^2_{SM}$ & {\color{white}{{--}}}& I & $\chi^2_{I}$ & {\color{white}{{--}}} & II &  $\chi^2_{II}$ \\
 \hline
     {\color{blue}{$\sigma_{\rm had}^0$   [nb]} }     & 41.541(37) &&  41.481 &  {\color{blue}{{\bf 2.6}}} &&41.541 &  {\color{blue}{{\bf 0.0}}} & & 41.532 & {\color{blue}{{\bf 0.1}}}\\
    $R_b (-2)$            & 0.2142(27)		     &&  0.2150 & 0.1 && 0.2157 & 0.3 && 0.2161 & 0.5\\
       $R_b^0$            &  0.21629(66)	   &&  0.21580 & 0.6 && 0.21693 & 1.0  && 0.21676 & 0.5\\
       $R_b (+2)$            &  0.2177(24)	   && 0.2155 & 0.8& & 0.2179 & 0.0 &&  0.2185 & 0.1\\
       {\color{blue}{$A_{FB}^b (-2)$  }}          & 0.0560(66) &   &  0.0638 &{\color{blue}{{\bf 1.4}}}&& 0.0581 & {\color{blue}{{\bf 0.1}}} && 0.0583 & {\color{blue}{{\bf 0.1}}}\\
       {\color{blue}{$A_{FB}^b ({\rm pk})$  }}          &  0.0982(17)	  & &  0.1014  &   {\color{blue}{{\bf 3.5}}} && 0.0980 & {\color{blue}{{\bf 0.0}}}  & &0.0971 &  {\color{blue}{{\bf 0.4}}} \\
       {\color{blue}{$A_{FB}^b (+2)$   }}         &  0.1125(55)	   &&  0.1255 & {\color{blue}{{\bf 5.6}}}&& 0.1133 & {\color{blue}{{\bf 0.0}}} & &0.1139 & {\color{blue}{{\bf 0.1}}} \\
      $A_b$       &  0.924(20) &&  0.935 & 0.3 && 0.921 & 0.0  &&  0.923& 0.0\\
      $R_e^0$       &  20.804(50) &&  20.737 & 1.8 && 20.772 & 0.4  && 20.759 & 0.8\\
                  $A_{FB}^{0,e}$       &  0.0145(25) &&  0.0165  & 0.7 && 0.0176 & 1.6  && 0.0167 & 0.7\\
      {\color{blue}{$A_e{(\rm LR-had.)}$ }}      &  0.15138(216) && 0.14739  & {\color{blue}{{\bf 3.4}}} && 0.15047 & {\color{blue}{{\bf 0.2}}} &&  0.14849& {\color{blue}{{\bf 1.8}}}\\
      $A_e{(\rm LR-lept.)}$      &  0.1544(60) && 0.1473  & 1.4 && 0.1473 & 1.4  && 0.1474 & 1.4\\
        \hline
      total  $\chi^2$ &  &  &   &  24.6 & && 6.76 &  &&9.99  \\
            \hline
  \hline
      \end{tabular}
   \label{tab:fit}
\end{table}

Besides quantities included in the $\chi^2$ and given in  Table~\ref{tab:fit} we check all other electroweak data on and near the Z-pole,  and above and below the $Z$-pole. 
While b-quark quantities were measured at three energies near the $Z$-pole, the total hadronic cross section was measured also at $\pm 1,3$ GeV
as discussed in the previous section.
The measurement at $+1$ GeV roughly coincides with the $Z'$-peak where the deviation from the SM would be the largest. The experimental error in  $\sigma_{\rm had}$  at  $+1$ GeV from the peak is $\sim 1\%$ for each LEP experiment and thus the $Z'$-peak contributes only a  fraction of the error bar. 

At energies above the $Z$-pole, the  $A_{FB}^b$ in the $Z'$ model basically coincides with the SM prediction while
$R_b$  fits data better than the SM, see Fig.~\ref{fig:NP}, with   $\chi^2 = 4.9$ for 10 data points compared to the SM which has   $\chi^2 = 7.3$ (the average discrepancy with respect to the SM prediction for $R_b$ is $-2.1\sigma$)~\cite{Alcaraz:2006mx}. At energies below the $Z$-pole the $Z'$ leads only to negligible differences from the SM predictions compared to sensitivities of current experiments.  


The quantities related to other charged leptons and quarks are not directly affected by $Z'$ and the predictions are essentially identical to predictions of the SM~\cite{Nakamura:2010zzi}.   
For example, the LEP 1 average of leptonic asymmetry assuming lepton universality, $A_l = 0.1481 \pm 0.0027$,
agrees very well with the SM prediction and would be only negligibly altered by the $Z'$ with couplings corresponding to the best fit (the prediction is  the same as for $A_e{(\rm LR-lept.)}$ given in Table~\ref{tab:fit}).

\begin{figure}[t!] 
   \centering
   \includegraphics[width=2.6in]{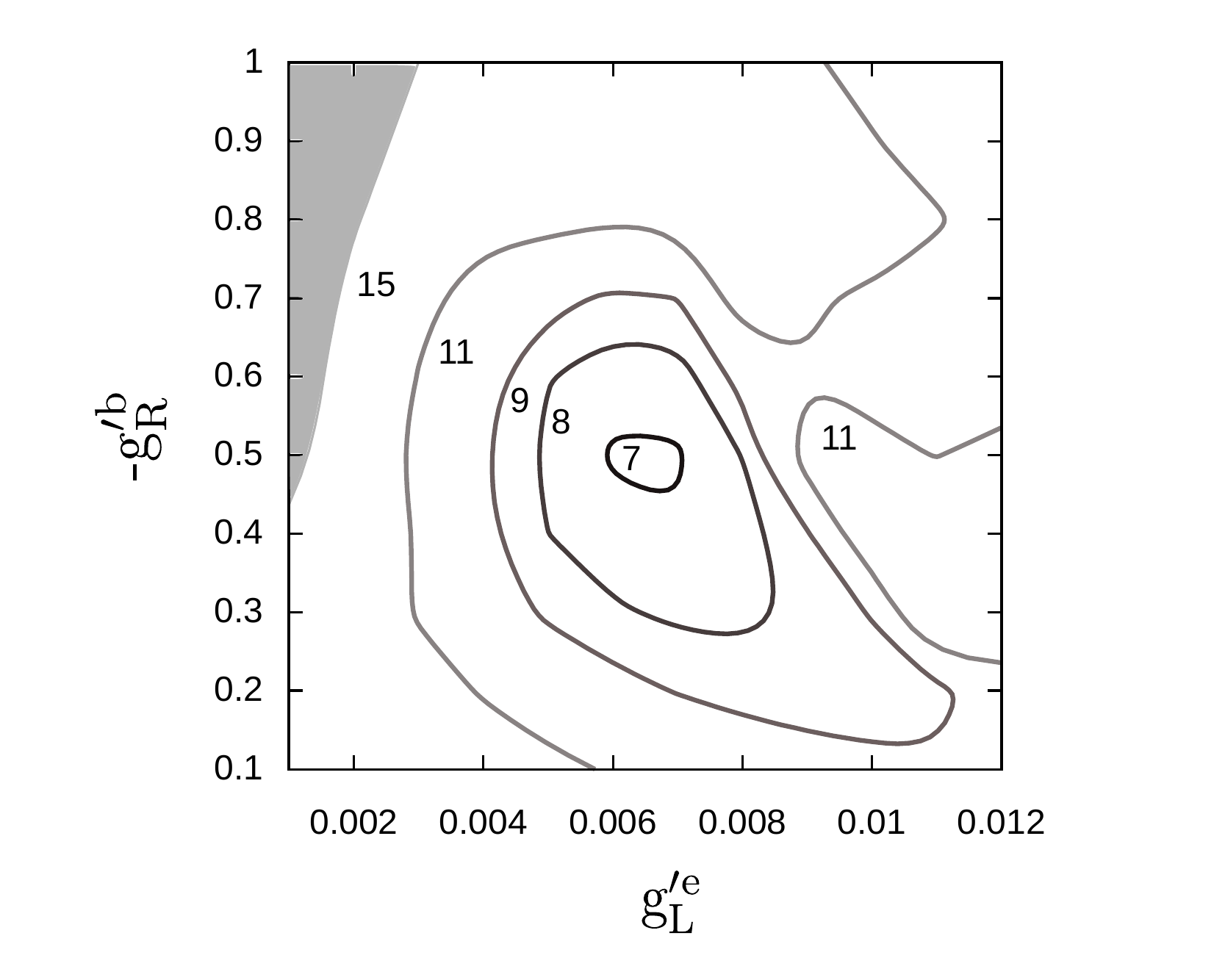}      
   \caption{Contours of constant $\chi^2$. Couplings $g'^{e}_R$ and $g'^{b}_L$ are free parameters. The shaded region corresponds to $\chi^2 > 15$. }
   \label{fig:chi2}
\end{figure}

\begin{figure}[h] 
   \centering
\includegraphics[width=4.5in]{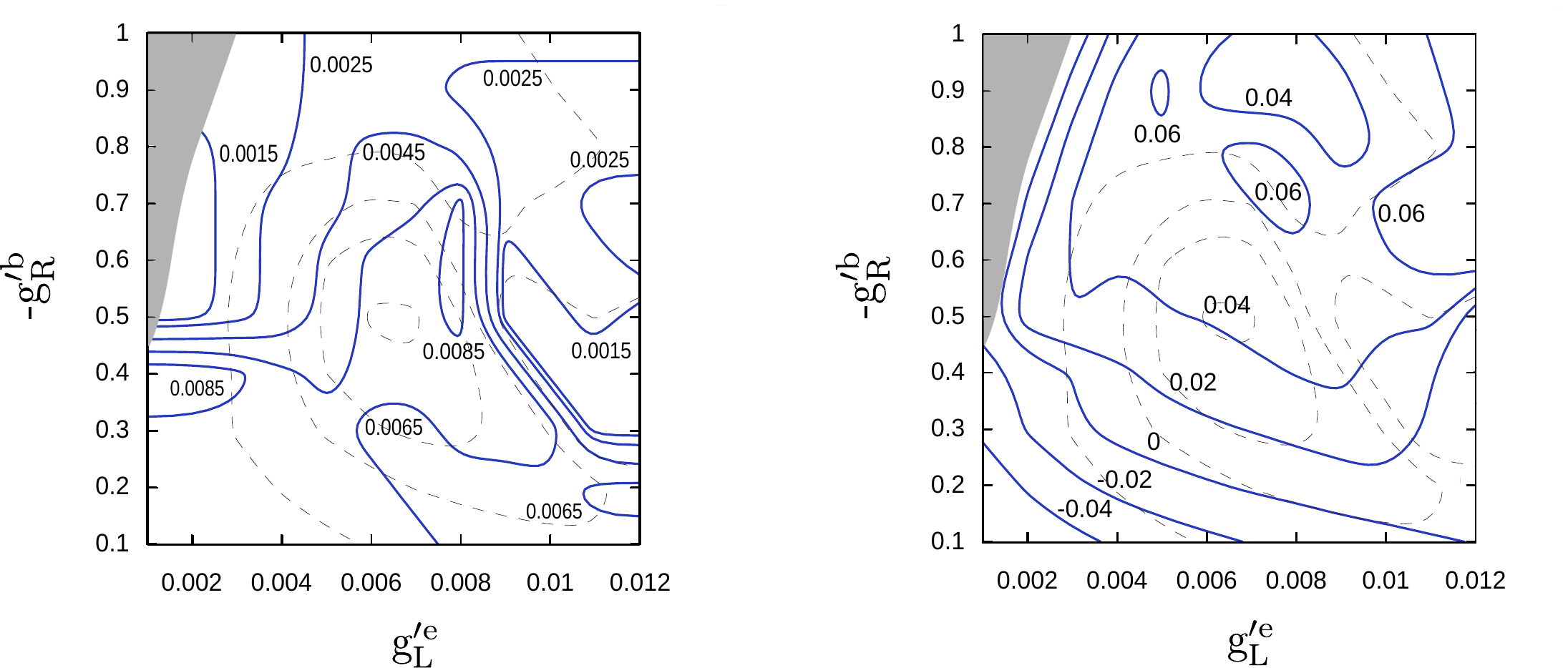}
   \caption{Contours of constant $g'^{e}_R$ (Left) and $g'^{b}_L$  (Right) from the best fit in the $g'^e_L - g'^b_R$ plane with $\chi^2$ contours from Fig.~\ref{fig:chi2} overlayed.}
   \label{fig:eR_and_bL}
\end{figure}

\begin{figure}[h] 
   \centering
   \includegraphics[width=4.5in]{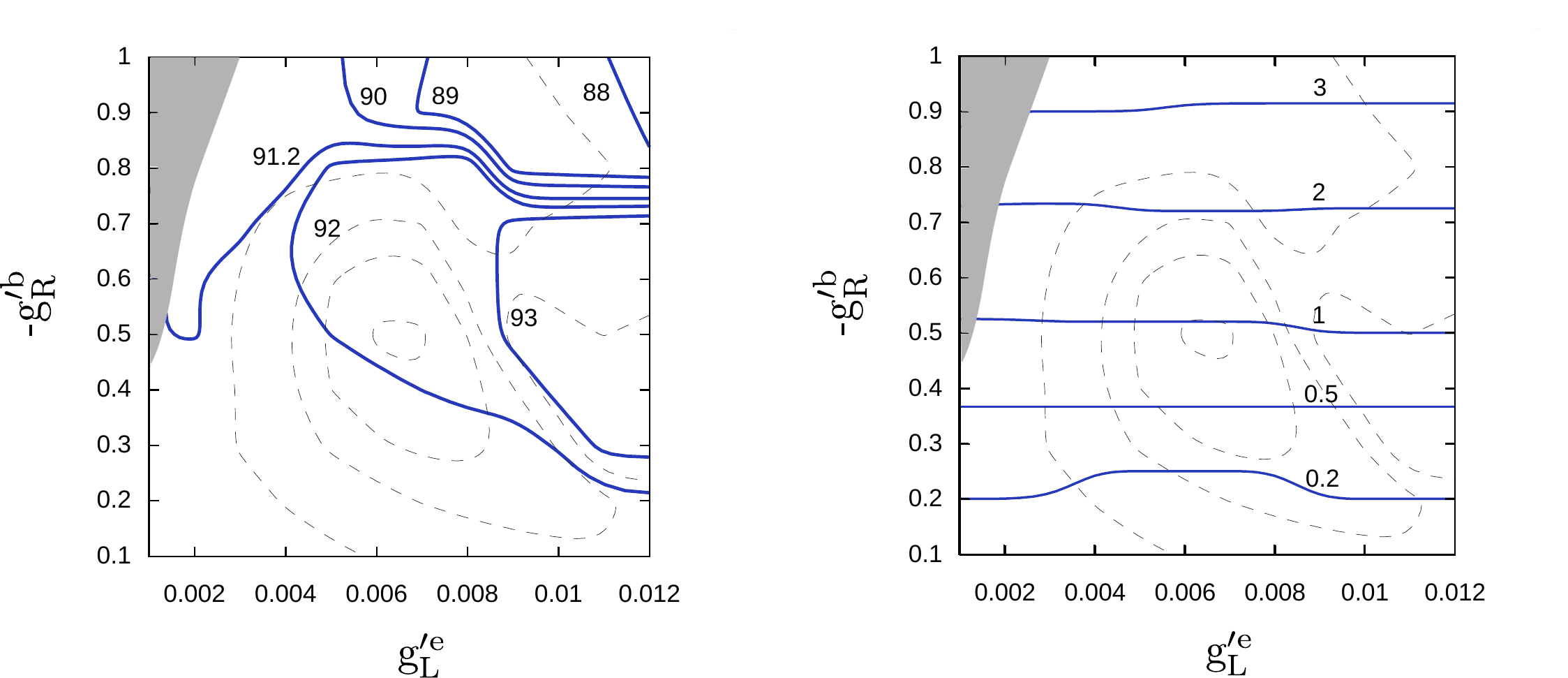}
   \caption{Contours of constant  $m_{Z'}$ (GeV) (Left) and the width of $Z'$ (GeV) (Right) with $\chi^2$ contours from Fig.~\ref{fig:chi2} overlayed.}
   \label{fig:mzp_and_width}
\end{figure}

The $\chi^2$ is not very sensitive to exact values of couplings. As can be seen from contours of constant $\chi^2$ in $g'^e_L - g'^b_R$ plane given in Fig.~\ref{fig:chi2}, a significantly better fit compared to the standard model can be achieved in a  large range of couplings. 
Contours  of the other two couplings, $g'^e_R$ and $g'^b_L $, corresponding to the best fit in the $g'^e_L - g'^b_R$ plane are given in Fig.~\ref{fig:eR_and_bL}, and contours of constant $m_{Z'}$ and the width of $Z'$ determined from its couplings to the electron and the bottom quark are given in Fig.~\ref{fig:mzp_and_width}.
In these and following plots the $\chi^2$ contours from Fig.~\ref{fig:chi2} are overlayed for easy guidance of the fit quality.

\begin{figure}[t!] 
   \centering
         \includegraphics[width=6.5in]{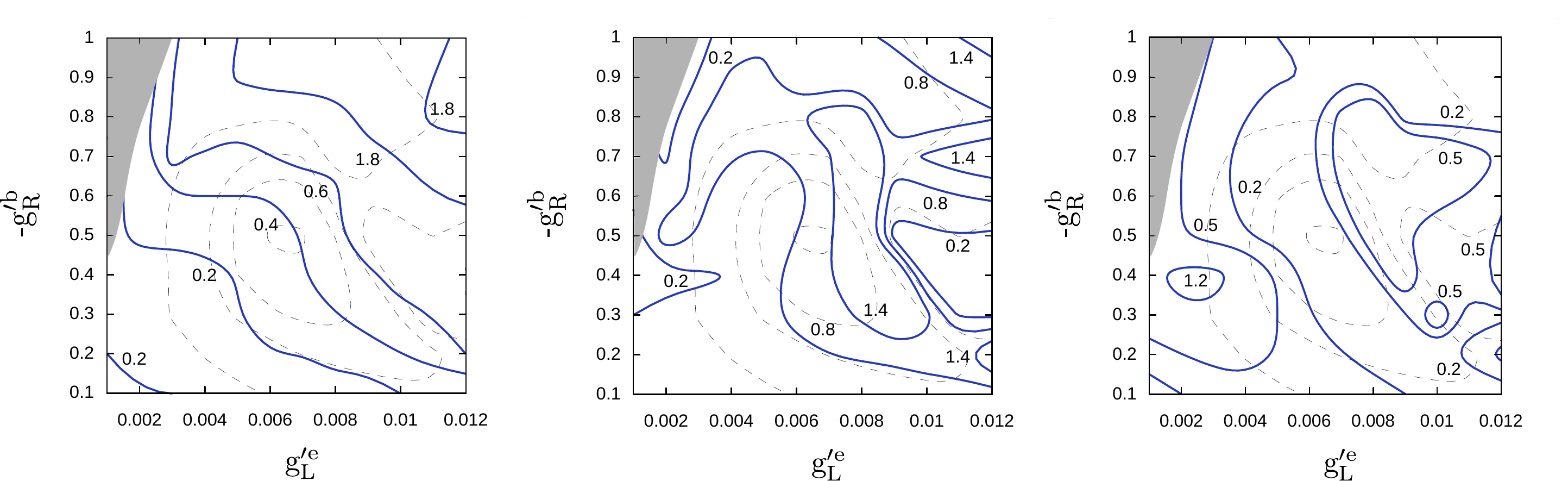}      
         \caption{Contours of constant contribution to $\chi^2$ from $R_b(-2)$ (Left), $R^0_b$ (Middle), and $R_b(+2)$ (Right) with $\chi^2$ contours from Fig.~\ref{fig:chi2} overlayed.}
   \label{fig:Rb}
\end{figure}

\begin{figure}[h!] 
   \centering
         \includegraphics[width=6.5in]{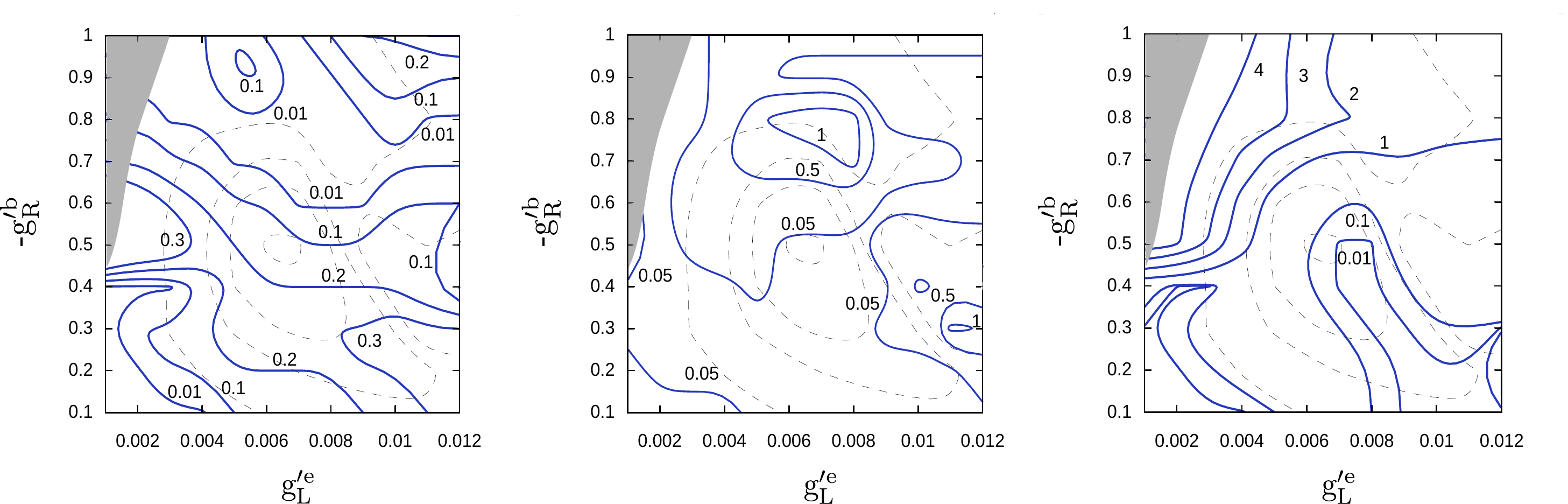}      
   \caption{Contours of constant contribution to $\chi^2$ from $A_{FB}^b(-2)$ (Left),  $A_{FB}^b(\rm pk)$ (Middle), and  $A_{FB}^b(+2)$ (Right)  with $\chi^2$ contours from Fig.~\ref{fig:chi2} overlayed.}
   \label{fig:AFBb}
\end{figure}

\begin{figure}[h!] 
   \centering
         \includegraphics[width=6.5in]{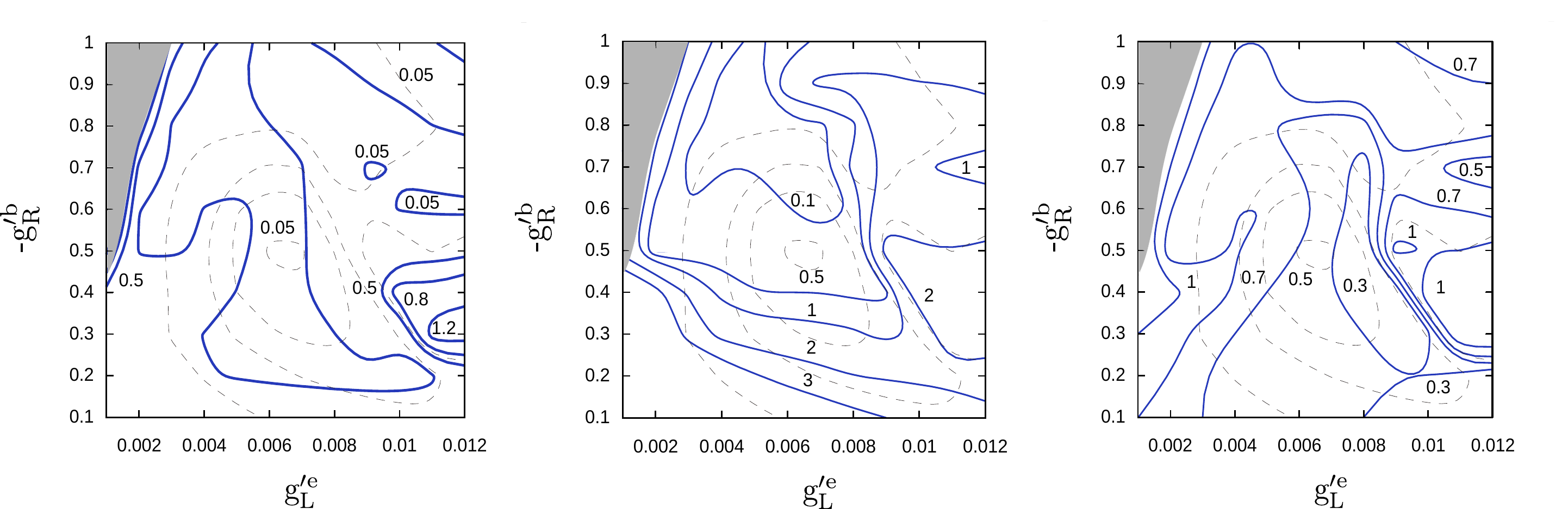}      
   \caption{Contours of constant contribution to $\chi^2$ from $\sigma^0_{\rm had}$ (Left),  $A_e(\rm LR-had.)$ (Middle), and $R_e^0$ (Right) with $\chi^2$ contours from Fig.~\ref{fig:chi2} overlayed.}
   \label{fig:sig_and_Aehad_and_Re}
\end{figure}

Partial contributions to $\chi^2$ from each observable are given in Figs.~\ref{fig:Rb} -- \ref{fig:sig_and_Aehad_and_Re}. Those from Table~\ref{tab:fit} that are not plotted, namely $A_b$,  $A_{FB}^{0,e}$,  and  $A_e{(\rm LR-lept.)}$, vary negligibly with varying the couplings. From these plots we clearly see that the main drivers toward the region of the best fit are 
$A_{FB}^b (+2)$, given in Fig.~\ref{fig:AFBb} (Right), disfavoring the upper-left region of couplings in the $g'^e_L - g'^b_R$ plane, and $A_e{(\rm LR-had.)}$, given in Fig.~\ref{fig:sig_and_Aehad_and_Re} (Middle), disfavoring the lower-right region. The three measurements of $R_b$  further constrain the allowed region of couplings approximately along the diagonal, see Fig.~\ref{fig:Rb}. The $\sigma^0_{\rm had}$, given in Fig.~\ref{fig:sig_and_Aehad_and_Re} (Left), fits close to the central value in a large range of couplings, and finally, $R_e^0$ prefers the central and lower region in the 
$g'^e_L - g'^b_R$ plane, see Fig.~\ref{fig:sig_and_Aehad_and_Re} (Right).


\begin{figure}[t] 
   \centering
   \includegraphics[width=3.5in]{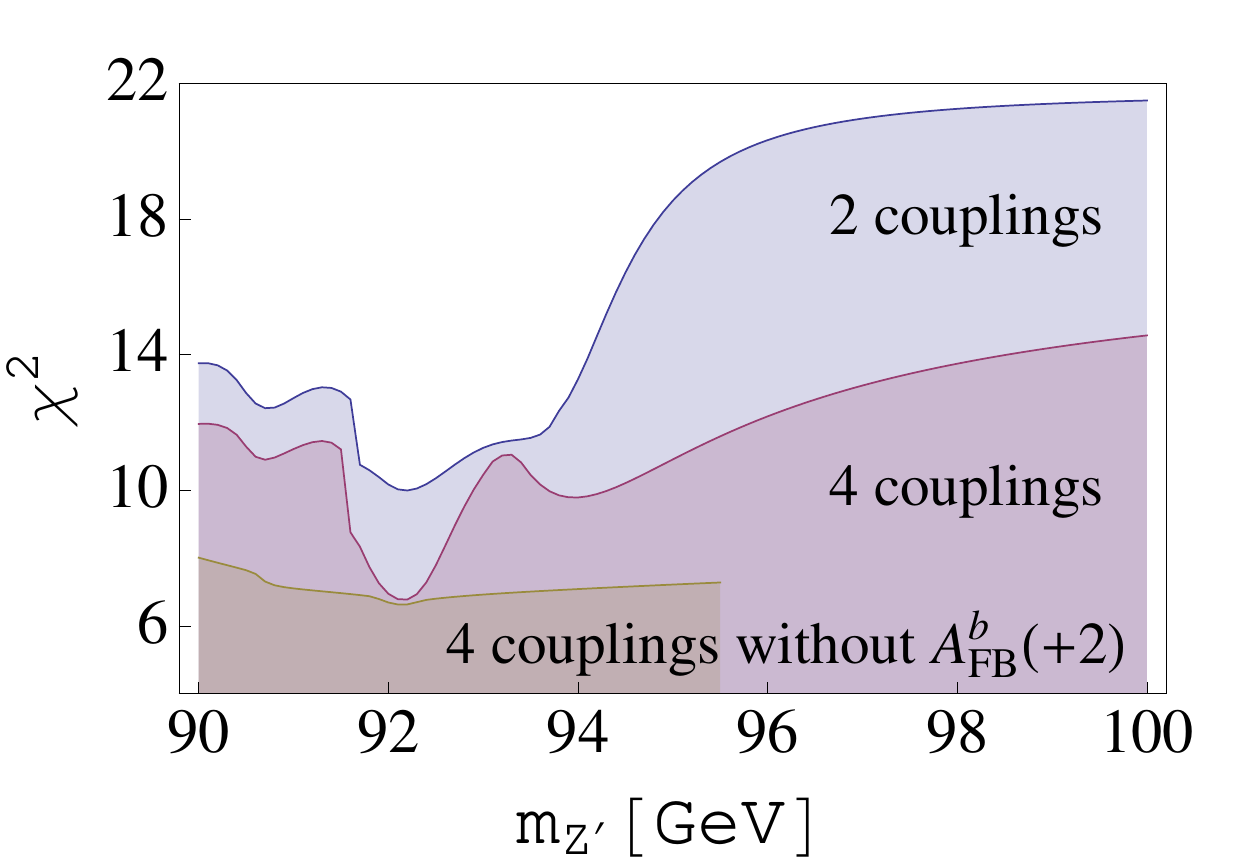}      
   \caption{Minimum of  $\chi^2$ as a function of $m_{Z'}$ for the fit with all for couplings being free parameters (middle line), and for the fit with only two couplings being free parameters, assuming $g'^{e}_R = g'^{b}_L = 0$ (top line). In addition, a fit without $A_{FB}^b(+2)$ in the $\chi^2$ function near the region of the best fit is shown (bottom line) demonstrating that the best fit value of $m_{Z'}$ is mainly driven by the $+2$ GeV measurement of the $A_{FB}^b$.}
   \label{fig:chi2_vs_mZ'}
\end{figure}

We have seen in Fig.~\ref{fig:mzp_and_width} (Left) that the minimum of $\chi^2$ prefers  $m_{Z'}$ close to the mass of the $Z$ boson with the best fit 
requiring $Z'$ just $\sim 1$ GeV heavier than the Z boson,  $m_{Z'} = 92.2$ GeV. However, much better fit compared to the standard model can be obtained even for somewhat heavier $Z'$.
Minimum of  $\chi^2$ as a function of $m_{Z'}$ for the fit with all for couplings being free parameters and for the fit with only two couplings being free parameters is plotted in Fig.~\ref{fig:chi2_vs_mZ'}. In the same figure, we also show the best fit with $A_{FB}^b(+2)$ removed from the $\chi^2$ function near the region of the best fit. This $\chi^2$ function is almost flat which demonstrates that the best fit value of $m_{Z'}$ is mainly driven by the $+2$ GeV measurement of the $A_{FB}^b$.

\begin{figure}[t] 
   \centering
   \includegraphics[width=4.5in]{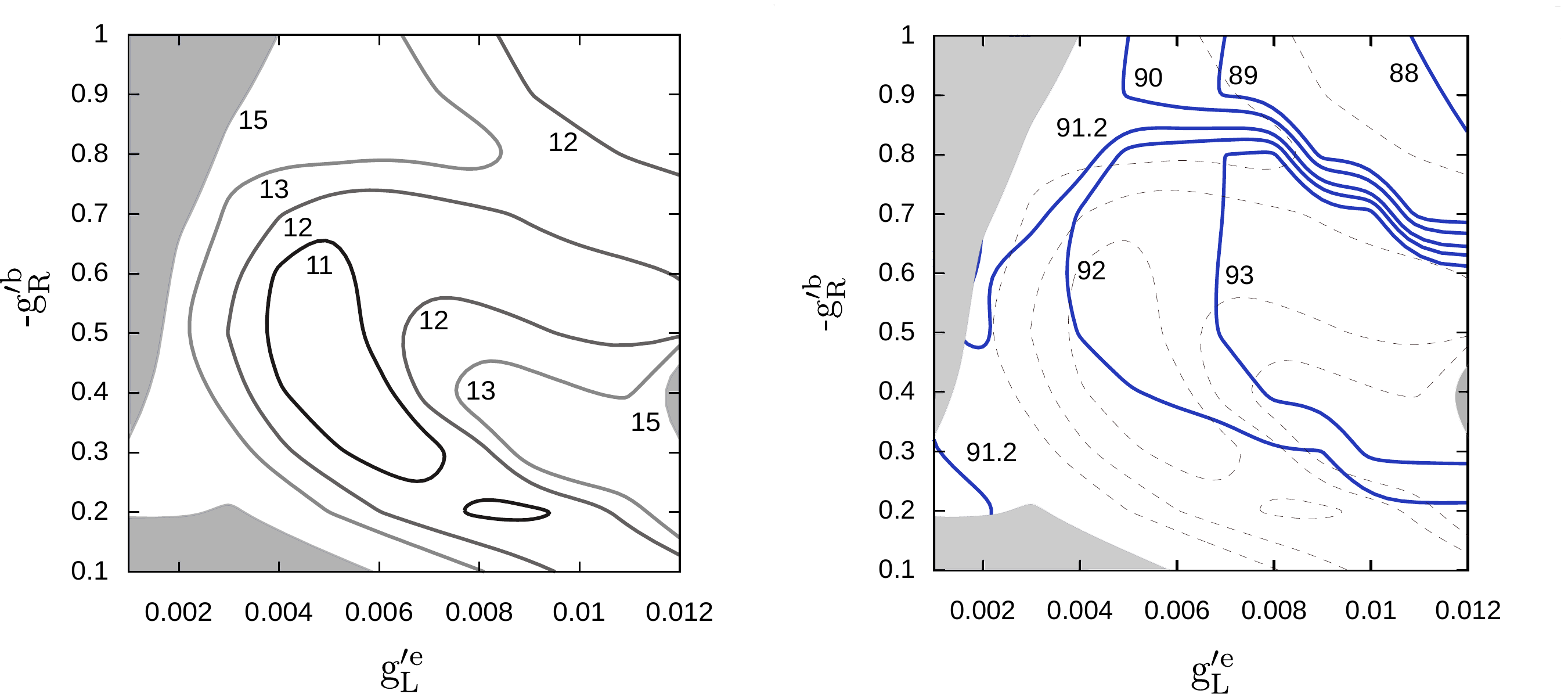}
   \caption{Contours of constant $\chi^2$ (Left) and $m_{Z'}$ (GeV) (solid lines) with $\chi^2$ contours (dashed lines) (Right) with $g'^{e}_R = g'^{b}_L = 0$.  The shaded region corresponds to $\chi^2 > 15$. }
   \label{fig:2c_chi2_and_mzp}
\end{figure}

\begin{figure}[t] 
   \centering
    \includegraphics[width=4.5in]{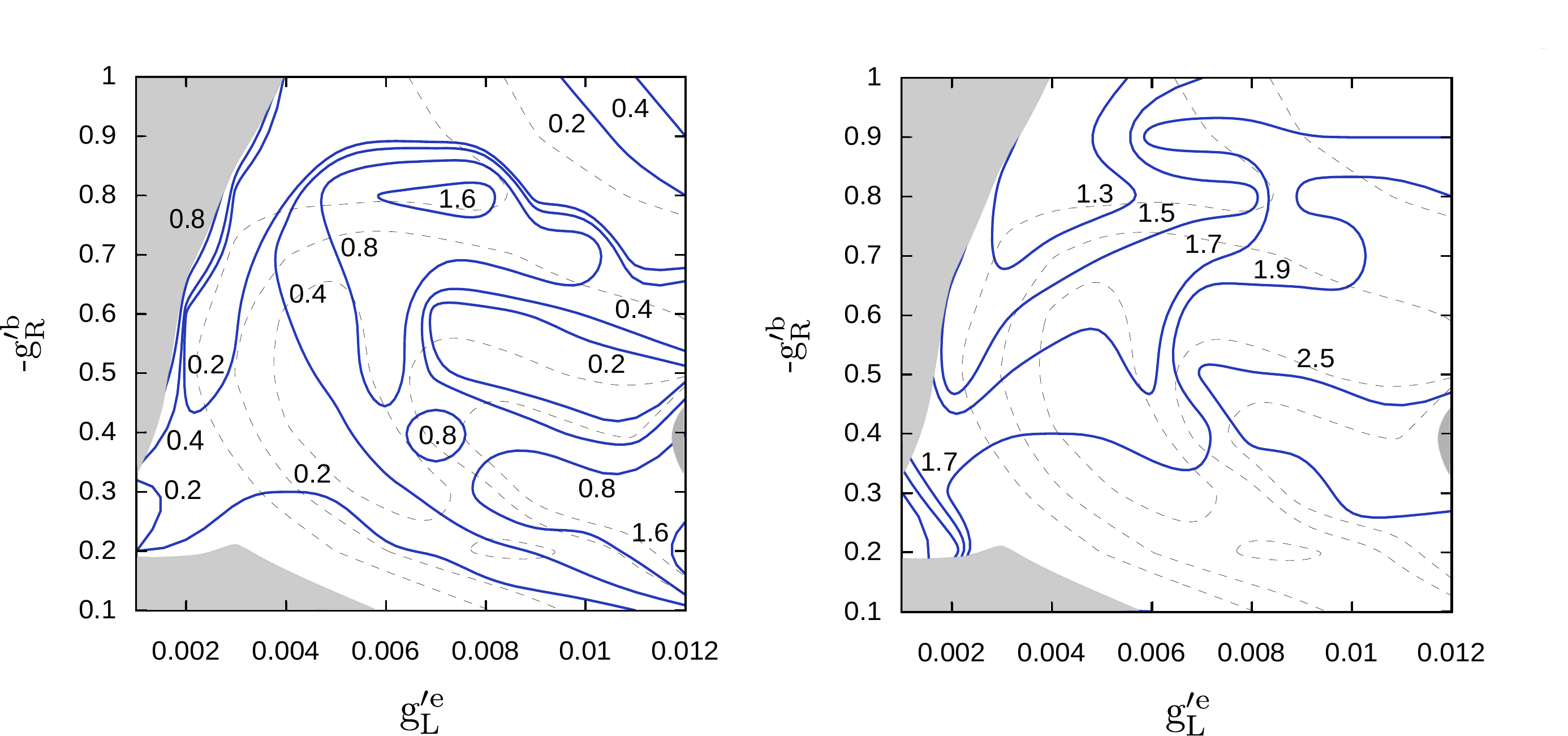}
   \caption{Contours of constant contribution to $\chi^2$ from $A_{FB}^b(\rm pk)$ (Left) and $A_e(\rm LR-had.)$ (Right) with $g'^{e}_R = g'^{b}_L = 0$. The $\chi^2$ contours from Fig.~\ref{fig:2c_chi2_and_mzp} are overlayed.}
   \label{fig:2c_AFBb_and_Aehad}
\end{figure}

Finally, let us comment on the case with only 2 allowed couplings when moving away from the best fit presented in Table~\ref{tab:fit}.
Contours of constant $\chi^2$ in $g'^e_L - g'^b_R$ plane in the case when only these two couplings are allowed, and thus $g'^{e}_R = g'^{b}_L = 0$, are given in Fig.~\ref{fig:2c_chi2_and_mzp} (Left). Preferred region of these couplings is very similar to the case with all four couplings allowed, see Fig.~\ref{fig:chi2}, which demonstrates that   $g'^e_L$ and $g'^b_R$ are the relevant couplings responsible for dramatic improvement of the fit compared to the standard model. Contours of constant $m_{Z'}$, given in Fig.~\ref{fig:2c_chi2_and_mzp} (Right), also closely resemble those of the four coupling fit, see Fig.~\ref{fig:mzp_and_width} (Left). The main difference from  the previous fit with all four couplings allowed and the reason for somewhat worse $\chi^2$ are $A_{FB}^b(\rm pk)$ and  $A_e(\rm LR-had.)$  given in Fig.~\ref{fig:2c_AFBb_and_Aehad}. Contributions to $\chi^2$ from other observables is very similar to the previous case with all four couplings.

\section{A possible model and its consequences}

In order to have large enough contribution to $Z$-pole observables without significantly modifying above the $Z$-pole measurements the mass of the $Z'$ should be within few GeV from the $Z$ mass. Couplings that are required are: $g'^{e}_L \simeq 0.005$ and $g'^{b}_R \simeq -0.5$. 
Additional small $g'^{e}_R$ and $g'^{b}_L$ further improve the fit to Z-pole data but are not required. Their presence however expands regions of 
$g'^{e}_L$ and $g'^{b}_R$ where a good fit is achieved.

The required couplings of standard model fermions to $Z'$  do not follow the usual pattern expected from new gauge interactions. 
A simple framework to generate arbitrary couplings of standard model fermions to a $Z'$ while preserving Yukawa interactions and keeping the model anomaly free was recently discussed in Ref.~\cite{Fox:2011qd}. In this framework the couplings of standard model fields to $Z'$ are generated effectively  through mixing with extra vector-like fermion pairs.  We will follow this direction and customize it for our purposes.

\begin{table}[h]
    \caption{Quantum numbers of relevant standard model and extra vector-like particles.}
    
\begin{tabular}{l r r r r r r r r r r}
 \hline
  \hline
   & $q_L$ & $d_R$ & $l_L$ & $e_R$ & $H$ & $D_L$ & $D_R$ & $L_L$ & $L_R$ & $ \Phi$ \\
 \hline
SU(3)$_{\text C}$ & {\bf 3} & {\bf 3}  &  {\bf 1} &  {\bf 1} & {\bf 1} & {\bf 3} & {\bf 3} &  {\bf 1} &  {\bf 1} & {\bf 1} \\
SU(2)$_{\text L}$ & {\bf 2} & {\bf 1} & {\bf 2} & {\bf 1}  & {\bf 2} & {\bf 1} & {\bf 1} &{\bf 2} & {\bf 2} &  {\bf 1} \\
U(1)$_{\text Y}$   & $\frac{1}{6}$ & -$\frac{1}{3}$ & -$\frac{1}{2}$ & -1 &$\frac{1}{2}$& -$\frac{1}{3}$ & -$\frac{1}{3}$ & -$\frac{1}{2}$ & -$\frac{1}{2}$ & 0 \\
U(1)'         & 0 &  0 & 0 & 0 & 0 & -1 & -1 & 1& 1& -1 \\
        \hline
  \hline
      \end{tabular}
   \label{tab:charges}
\end{table}

Let us start with adding a vector-like pair of fermions $D_L$ and $D_R$ charged under a $U(1)'$ where $D_R$ has the same quantum numbers under the standard model gauge symmetry as the $d_R$, see Table~\ref{tab:charges}. This charge assignment results in the following renormalizable terms in the lagrangian:
\begin{equation}
{\cal L} \supset -  \bar q_{Li} Y^d_{ij} d_{Rj} H - \bar D_L \lambda^d_k d_{R k} \Phi - \mu_D \bar D_L D_R + {\it h.c.}
\end{equation}
where the first term represents the usual standard model Yukawa couplings for down-type quarks (the sum over flavor indices is assumed). The second term contains Yukawa interactions of standard model quarks and the extra $D_L$ quark. The last term is the mass term for the vector-like pair. 
The vacuum expectation value of the scalar field $\Phi$ breaks the $U(1)'$ and generates mixing terms between $d_{Ri}$ and $D_R$. After spontaneous symmetry breaking the $4 \times 4$ mass matrix for down type quarks is given by:
\begin{eqnarray}
( \bar d_{Li}, \bar D_L) M_d  \begin{pmatrix}
 d_{Rj} \\
 D_R
\end{pmatrix}
= 
( \bar d_{Li}, \bar D_L)
\begin{pmatrix}
 Y^d_{ij} \langle H \rangle & 0 \\
 \lambda^d_{j} \langle \Phi \rangle & \mu_D 
\end{pmatrix}
\begin{pmatrix}
 d_{Rj} \\
 D_R
\end{pmatrix},
\end{eqnarray}
and it can be diagonalized by a bi-unitary transformation, $U^\dagger_L M_d U_R$, which defines the mass eigenstate basis. 
However, before doing that,  it is instructive to  
change the basis by a unitary transformation,  $d_{R} \to V_R d_R$,  $d_{L} \to V_L d_L$, which diagonalizes the standard model Yukawa couplings, $Y^d$. The  mass matrix becomes:
\begin{eqnarray}
\begin{pmatrix}
 (V^\dagger_{L} Y^d V_{R} )_{ij}
 \langle H \rangle & 0 \\ \lambda^d_{n} V_{Rnj} \langle \Phi \rangle & \mu_D 
\end{pmatrix}
%
=
\, \mu_D
\begin{pmatrix}
 \beta_j \delta_{ij}& 0 \\
 \alpha_j & 1
\end{pmatrix},
\end{eqnarray}
where
\begin{equation}
\alpha_j = \frac{  \lambda^d_{n} V_{Rnj} \langle \Phi \rangle}{\mu_D}, \quad  \quad {\rm and} \quad  \quad
\beta_j = \frac{(V^\dagger_{L} Y^d V_{R})_{jj} \langle H \rangle}{\mu_D}.
\end{equation}
From this form of the mass matrix  we can see that in any theory of flavor that determines the structure of Yukawa matrices for standard model fermions, in this case only $Y_d$, but allows arbitrary $\lambda^d_i$ couplings, these can be chosen so that $\alpha_1 = \alpha_2 = 0$ and only $\alpha_3$ is non-zero. This corresponds to the situation when $\lambda^d_i \propto V_{R3i}^*$, or in the basis where standard model Yukawa couplings are diagonal, it corresponds to $ \lambda^d_1 = \lambda^d_2 = 0$ and $\lambda^d_3 \equiv \lambda_b$ is non-zero.   This is the minimal scenario that does not modify standard model couplings of down and strange quarks.
In what follows we will focus on this scenario. 

\begin{figure}[t] 
   \centering
   \includegraphics[width=2.3in]{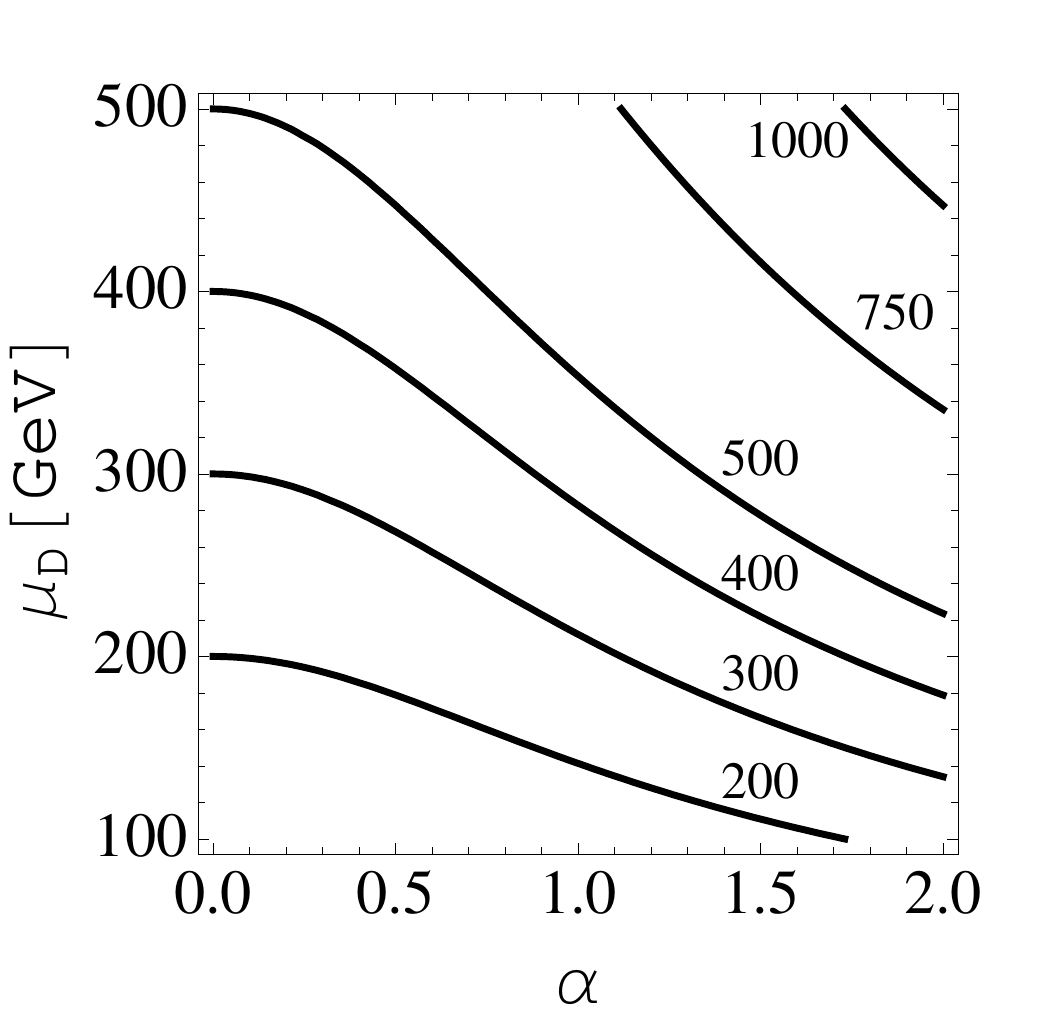}      
   \caption{The mass of the extra $D$ quark, $m_D$ [GeV], as a function of $\mu_D$ and $\alpha$.}
   \label{fig:mD}
\end{figure}

In the basis where standard model Yukawa couplings are diagonal, 
assuming $\lambda^d_i$ are such that $\alpha_1 = \alpha_2 = 0$,  the first two diagonal entries correspond to masses of the down and strange quarks:
\begin{equation}
m_{d,s} = \mu_D \,  \beta_{1,2}
\end{equation}
The lower $2 \times 2$ block can be diagonalized by a bi-unitary transformation (for simplicity we drop indices, $\alpha_3 \equiv \alpha$ and $\beta_3  \equiv \beta$):
\begin{eqnarray}
\mu_D \, U_L^\dagger 
\begin{pmatrix}
 \beta & 0 \\
\alpha & 1 
\end{pmatrix} U_R
=
\begin{pmatrix}
 m_b & 0 \\
0 & m_D 
\end{pmatrix} ,
\end{eqnarray}
where we use the same names,  $U_{L,R}$, for matrices that diagonalize the lower $2\times 2$ block in the case 
$\alpha_{1,2} = 0$, as for the matrices  that diagonalize the general $4\times4$ matrix. We label their components  by 3 and 4 so that results are applicable to the general scenario with non-zero $\alpha_{1,2}$. 
The bottom quark mass and the mass of the extra heavy down-type quark are given by:
\begin{eqnarray}
m_b &\simeq& \mu_D \beta /\sqrt{1 + \alpha^2}, \label{eq:mb}\\
m_D &\simeq& \mu_D \sqrt{1 + \alpha^2},
\end{eqnarray}
where we assume $\beta \ll 1, \alpha $.
The mass of the $D$ quark as a function of $\mu_D$ and $\alpha$ is plotted in Fig.~\ref{fig:mD}.
The diagonalization matrices are approximately given by:
\begin{eqnarray}
U_L^\dagger \simeq  
\begin{pmatrix}
1 & - \alpha \beta \\
\alpha \beta & 1
\end{pmatrix}, \quad \quad \quad
U_R \simeq  \frac{1}{\sqrt{1+a^2}}
\begin{pmatrix}
1 & \alpha \\
- \alpha & 1
\end{pmatrix}.
\label{eq:U_L,R}
\end{eqnarray}

\subsection{Couplings of the $Z'$ boson}

\begin{figure}[t] 
   \centering
   \includegraphics[width=2.3in]{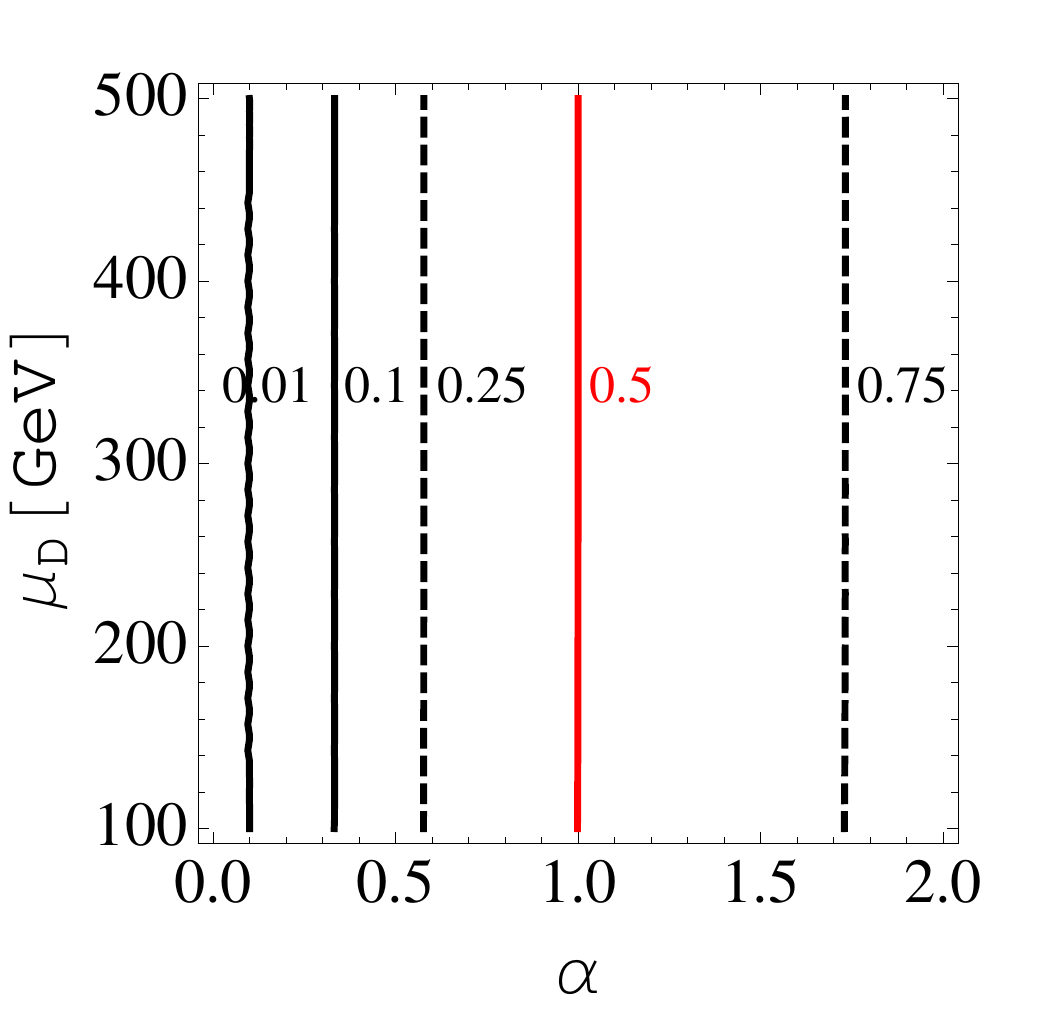}
   \includegraphics[width=2.3in]{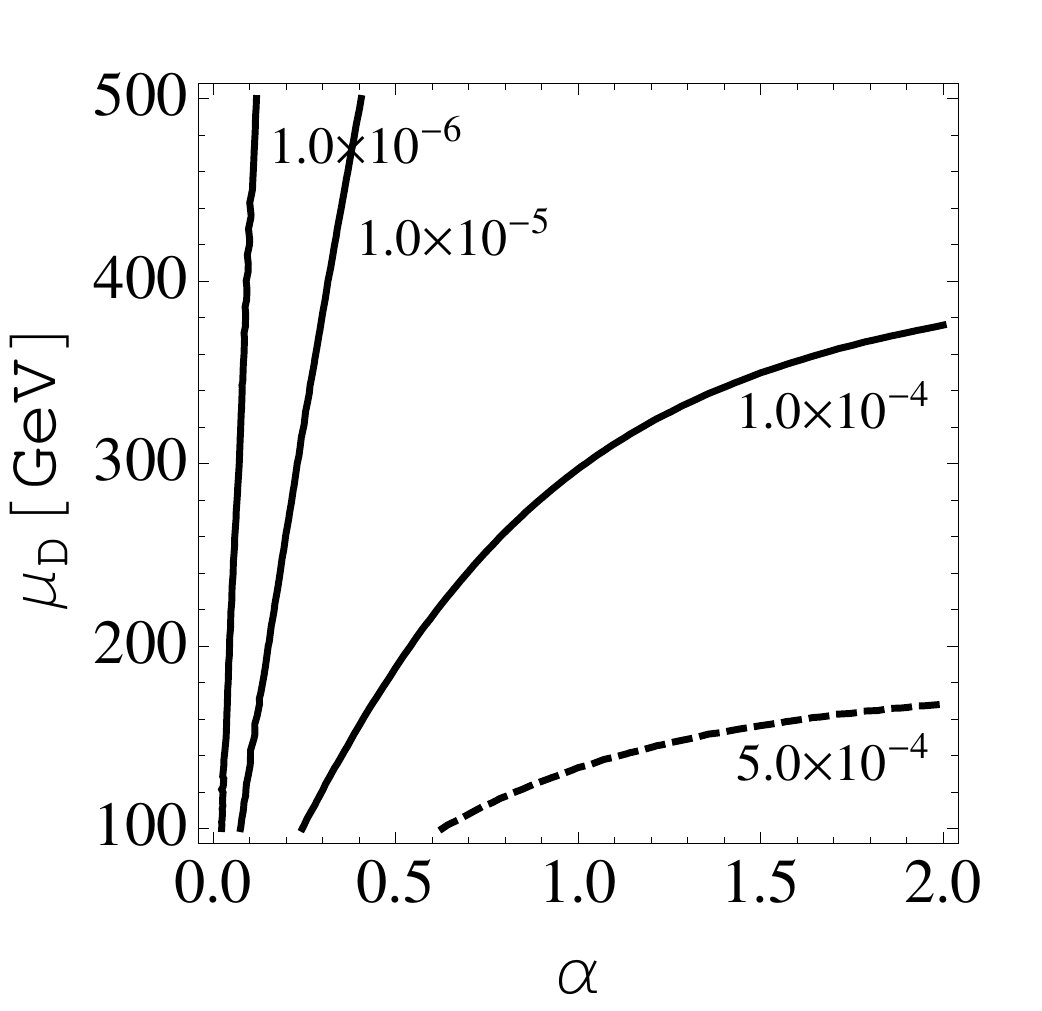}
   \caption{Contours of constant  -$g'^b_{R}$ (Left) and -$g'^b_{L}$ (Right) in  $\mu_D$  - $\alpha$ plane  for $g' = 1$.}
   \label{fig:g'b_mu_alpha}
\end{figure}

\begin{figure}[t] 
   \centering
   \includegraphics[width=2.3in]{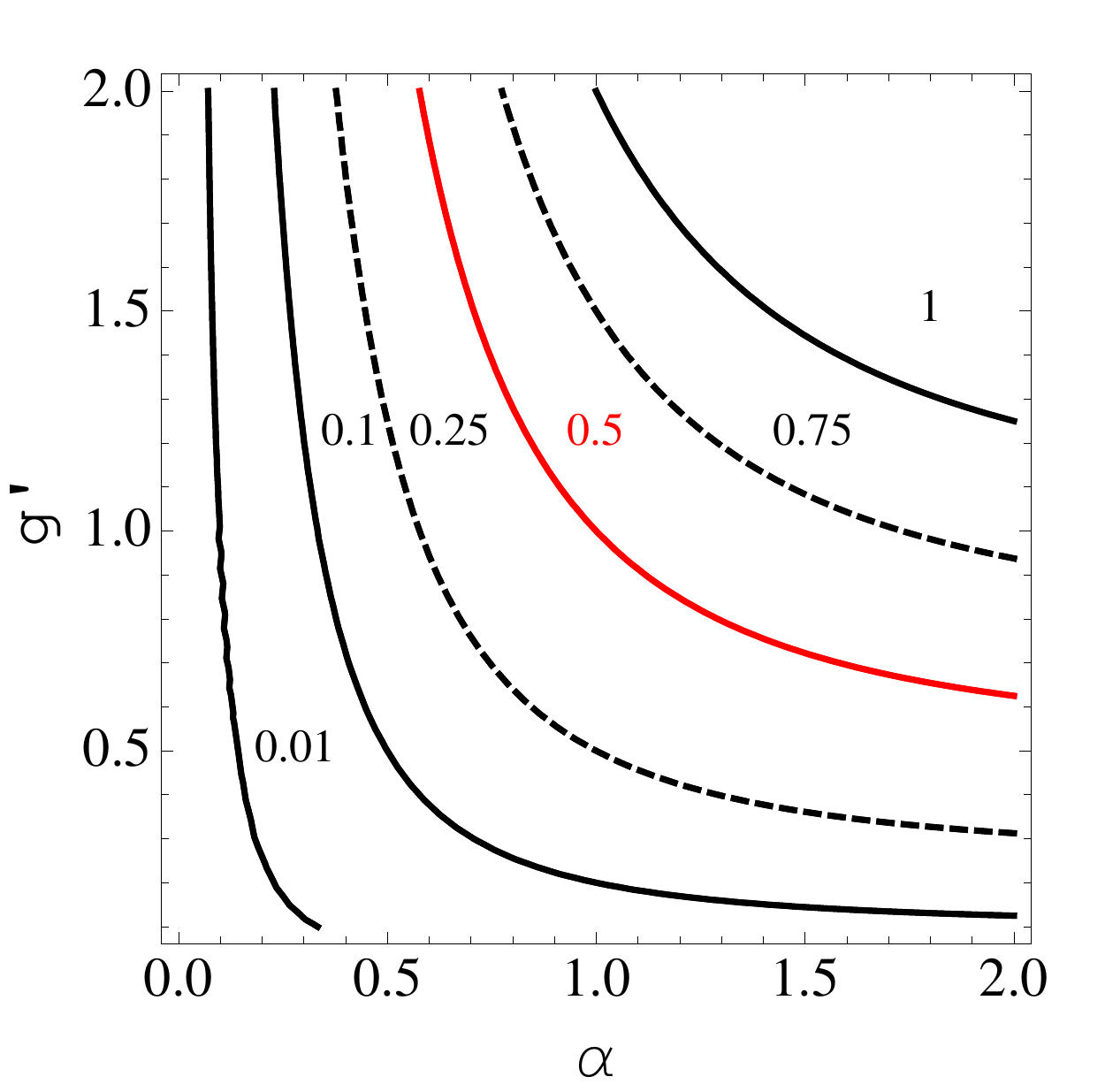}
   \includegraphics[width=2.3in]{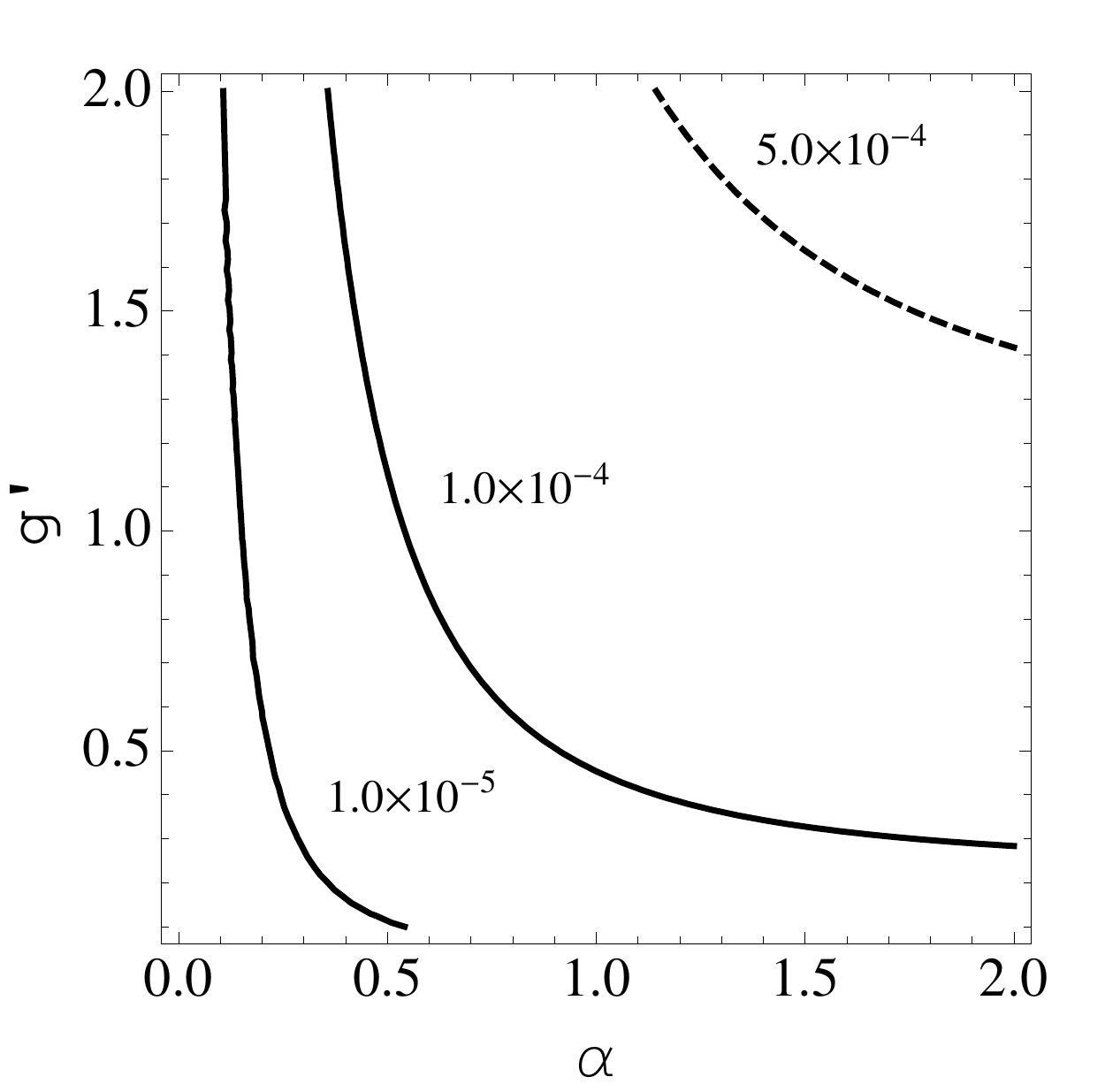}
   \caption{Contours of constant  -$g'^b_{R}$ (Left) and -$g'^b_{L}$ (Right) in  $g'$ - $\alpha$ plane for $\mu_D = 200$ GeV.}
   \label{fig:g'b_g'_alpha}
\end{figure}

Couplings of $Z'$ to down-type quarks (mass eigenstates) originate from the kinetic term of the extra vector-like pair:
\begin{eqnarray}
{\cal L}_{\text kin} \supset   \bar D_L i \slashed D' D_L +  \bar D_R i \slashed D' D_R \; = \; \bar {\hat d}_{L i} (U^\dagger_L)_{i 4} i \slashed D' (U_L)_{4 j} \hat d_{L j} +  \bar {\hat d}_{R i} (U^\dagger_R)_{i 4} i \slashed D' (U_R)_{4 j} \hat d_{R j},
\label{eq:kinD}
\end{eqnarray}
where the vectors of mass eigenstates are $\hat d_{R} = (d_R,s_R,b_R, \hat D_R)^T$ and similarly for $\hat d_{L}$.
The covariant derivative is given by:
\begin{eqnarray}
D'_\mu = D^{SM}_\mu - i g' Q' Z'_\mu,
 \end{eqnarray}
where $D^{SM}_\mu$  is the standard model covariant derivative:
\begin{eqnarray}
D^{SM}_\mu = \partial _\mu 
- i \frac{g}{\cos \theta_W} (T^3 - \sin ^2 \theta_W Q) Z_\mu - i e Q A_\mu,
\end{eqnarray}
and for simplicity we do not write  the $SU(3)_C$ interactions explicitly which are not modified by field redefinitions.
In the mass eigenstate basis, the Z' has in general both flavor diagonal and off-diagonal couplings to down-type quarks:
\begin{eqnarray}
g'^{f_i f_j}_R &= & -g' \; (U_{R}^\dagger)_{i 4} (U_{R})_{4 j} \\
g'^{f_i f_j}_L &= & -g' \; (U_{L}^\dagger)_{i 4} (U_{L})_{ 4 j} \label{eq:g'Lij},
\end{eqnarray}
where we used $Q'_D = -1$. For flavor diagonal couplings the expressions simplify to:
\begin{eqnarray}
g'^{f_i}_R &= & -g' \; |(U_{R})_{4 i} |^2 \\
g'^{f_i}_L &= & -g' \; |(U_{L})_{ 4 i}|^2. \label{eq:g'L}
\end{eqnarray}
In the case $\alpha_{1,2} = 0$, that we are focusing on, the first two generations do not have couplings to $Z'$ and only the bottom quark and the  $D$ quark couple to $Z'$ with couplings that can be obtained from Eq.~(\ref{eq:U_L,R}).
The $g'^b_{R,L}$ couplings as functions of $\mu_D$ and $\alpha$ assuming $g' = 1$ are given in Fig.~\ref{fig:g'b_mu_alpha},
and as  functions of $g'$ and $\alpha$,  for fixed $\mu_D = 200$ GeV, in Fig.~\ref{fig:g'b_g'_alpha}.
The $g'^b_{R}$ coupling is fully controlled by $\alpha$ and can be easily sizable. For $g' = 1$ it can be close to the value suggested by the best fit (highlighted in plots) for $\alpha \simeq1$.
The $g'^b_{L}$ coupling on the other hand is proportional to $\beta $ which is of order $m_b/\mu_D$, see Eq.~(\ref{eq:mb}), and thus it is very small. For the purposes of the fit to precision electroweak data it is effectively zero.

\subsection{Corrections to  neutral and charged currents}

All couplings of the photon and couplings of up-type quarks and right-handed down-type quarks to the Z boson are identical to standard model couplings. However, since $D_L$ is an SU(2)$_{\rm L}$ singlet,
 the couplings of left-handed down-type quarks to the Z boson are modified. They can be read out from kinetic terms of left handed fields (similar to Eq.~(\ref{eq:kinD}) but written for all four quarks):
 \begin{eqnarray}
g^{f_i f_j}_L &=& \frac{g}{\cos \theta_W} \displaystyle\sum\limits_{k=1}^4 (T^3_k - \sin ^2 \theta_W Q_d) (U^\dagger_L)_{ik} (U_L)_{kj}, 
\end{eqnarray}
where $T^3_k = -1/2$ for $k = 1,2,3$ and $0$ for $k=4$. Corrections to couplings of the $Z$  boson to  left-handed down-type quarks in the standard model ($i,j = 1,2,3$)  can be written as:
 \begin{eqnarray}
\delta g^{f_i f_j}_L &=& \frac{g}{2 \cos \theta_W}  (U^\dagger_L)_{i4} (U_L)_{4j}.
\label{eq:delgL}
\end{eqnarray}
In general, these corrections for the first two generations are tiny, since they are proportional to ratios of masses of  corresponding quarks and the heavy quark, $\delta g^{f_i f_j}_L \propto (m_i/\mu_D)(m_j/\mu_D)$.  In the case $\alpha_{1,2} = 0$, that we are focusing on, couplings of the first two generations  to $Z$ are not altered at all, and there are no flavor violating couplings. 

Comparing Eq.~(\ref{eq:delgL}) with Eqs.~(\ref{eq:g'Lij}) and (\ref{eq:g'L}) we  see that the change in a Z coupling is directly proportional to  corresponding $Z'$ coupling that is being generated.
For the correction to the left-handed bottom coupling we find:
 \begin{eqnarray}
\delta g^b_L &=& -\frac{g}{2 \cos \theta_W}  \frac{g'^b_L}{g'}.
\end{eqnarray}
and from the values of the ratio $g'^b_L/g'$ given in Fig.~\ref{fig:g'b_mu_alpha} we see that $\delta g^b_L$ is negligible.

\begin{figure}[t] 
   \centering
   \includegraphics[width=2.3in]{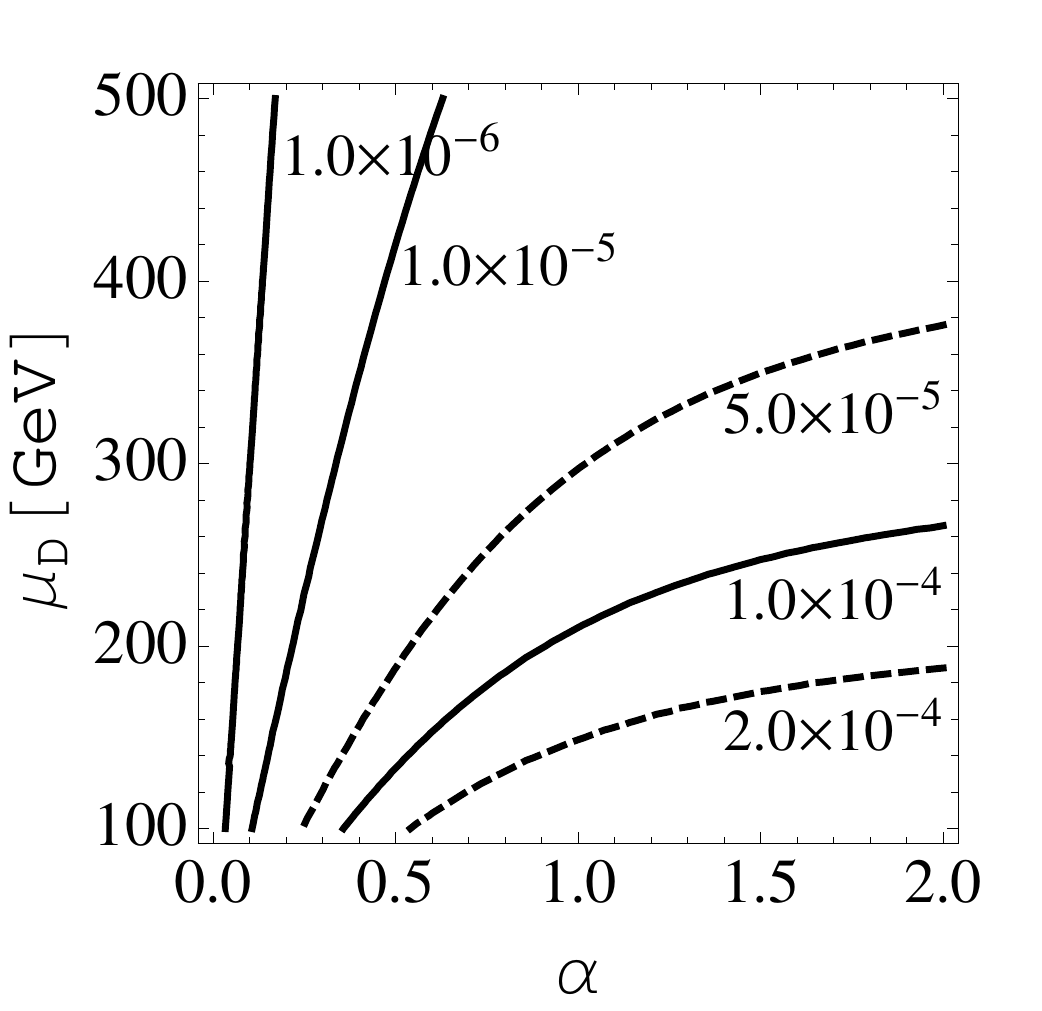}      
   \caption{Relative correction $\delta$ to the third column of the CKM matrix.}
   \label{fig:deltaCKM}
\end{figure}

The charge currents,
 \begin{eqnarray}
-\frac{g}{\sqrt{2}}  \bar  u_{Li} (V_{CKM})_{ij} \gamma^\mu d_{Lj} W^+_\mu + h.c.
\end{eqnarray}
get also modified by $d_{L} \to U_L \hat d_L$ which effectively leads to a modification of the CKM matrix:
 \begin{eqnarray}
(V_{CKM})_{ij} \to \displaystyle\sum\limits_{k=1}^3 (V_{CKM})_{ik} (U_L)_{kj}.
\end{eqnarray}
In the case $\alpha_{1,2} = 0$, only the third column of the matrix is modified:
 \begin{eqnarray}
(V_{CKM})_{ib} \to (V_{CKM})_{ib} (U_L)_{33}, \quad  \quad i = u,c,t.
\end{eqnarray}
It is convenient to define $\delta = 1 - (U_L)_{33}$ which represents the relative correction of the third column of the CKM matrix. It is plotted in Fig.~\ref{fig:deltaCKM}
and the values are far below current uncertainties in the CKM elements. 

\subsection{Exploring lagrangian parameters}

\begin{figure}[t] 
   \centering
   \includegraphics[width=2.8in]{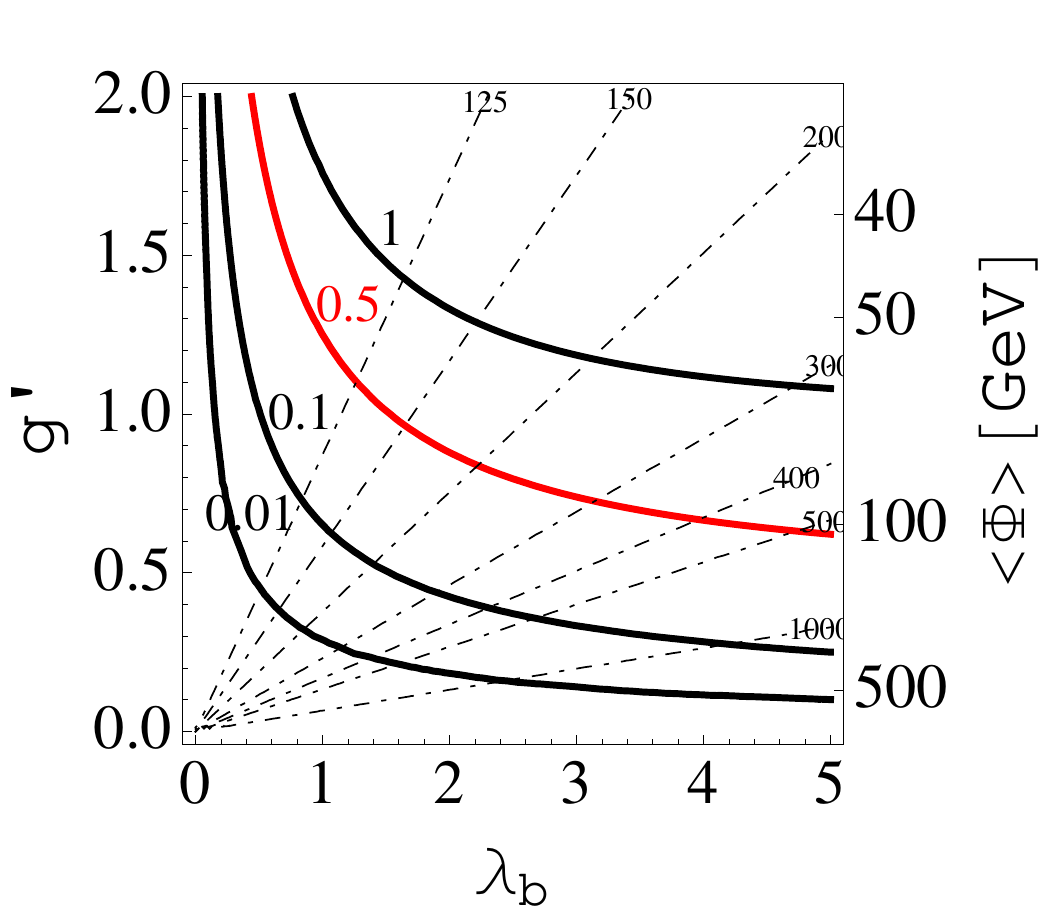}    \hspace{1.cm}
      \includegraphics[width=2.8in]{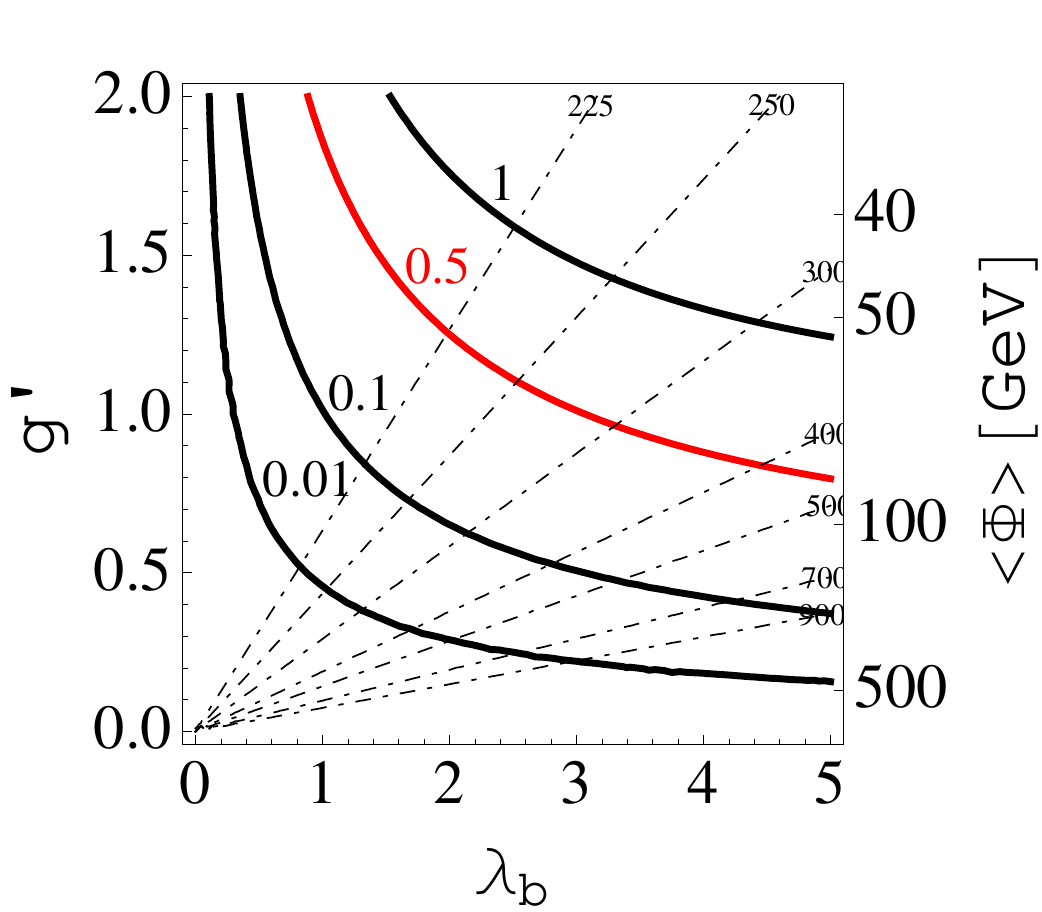}\\
       \includegraphics[width=2.8in]{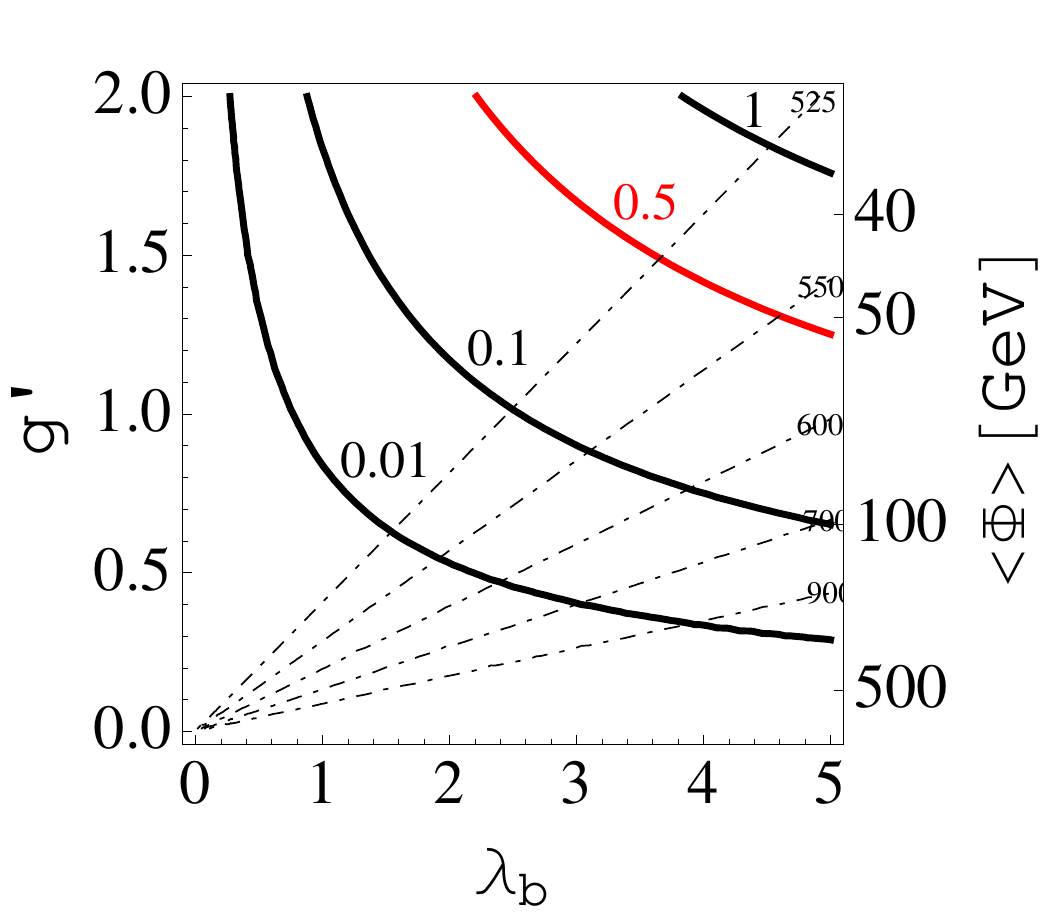}
   \caption{Contours of constant $g'^b_R$ (solid lines) in the $g'$ - $\lambda_b$ plane for  $\mu_D = 100$ GeV (Top Left), $200$ GeV (Top Right), and $500$ GeV (Bottom). The mass of the $Z'$ is fixed to the best fit value, $m_{Z'} = 92.2$ GeV, and the corresponding vev of the extra Higgs field is given on the right axis. The dashed lines represent contours of constant D quark mass, $m_D$ [GeV] (vertical line $\lambda_b = 0$ which is not shown would correspond to $m_D = \mu_D$).}
   \label{fig:g'lb}
\end{figure}

The model we have discussed so far is specified by  4 parameters: $g'$, $\lambda_b$, $\mu_D$, and the vacuum expectation value (vev) of the extra Higgs field that breaks the $U(1)'$ symmetry, $ \langle \Phi \rangle$. This vev  is responsible for generating the $Z'$ couplings to $b$ quark through mixing with $D$ (it is contained in  $\alpha$) and  also for the   mass term of the  $Z'$ boson:
\begin{eqnarray}
{\cal L}_\Phi \supset | D_\mu \Phi |^2 \supset  g'^2 \langle \Phi \rangle^2  Z'_\mu Z'^\mu,
\end{eqnarray}
\begin{eqnarray}
m_{Z'} &=& \sqrt{2} g' \langle \Phi \rangle. 
%
\end{eqnarray}
Equivalently, the  model is specified by $g'$, $m_{Z'}$, $\alpha$, and $\mu_D$,  although the fit to precision electroweak data depends only on two parameters: $m_{Z'}$ and $g'^b_R$. The fit strongly prefers mass of the $Z'$ close to the mass of the $Z$ boson and thus we can simply fix it to the best fit value $92.2$ GeV. In previous subsections we have explored the dependence of  $g'^b_R$   coupling  on $g'$,  $\alpha$, and $\mu_D$. It is however instructive to see what values of lagrangian parameters and $ \langle \Phi \rangle$ are required. Contours of constant $g'^b_R$ in the $g'$ - $\lambda_b$ plane for values of $\mu_D = 100$, 200, and 500 GeV are given in Fig.~\ref{fig:g'lb}. The corresponding vev of $\Phi$ is given on the right axis and the mass of the $D$ quark is overlayed. In order to obtain $g'^b_R \simeq 0.5$, as suggested by the best fit, while keeping $g'$ and $\lambda_b$ perturbative, the extra $D$ quark  should be fairly light, in a few hundred GeV range. However we should keep in mind that even $g'^b_R \simeq 0.1$ provides a significant improvement of the fit compared to the standard model, in which case the $D$ quark can be heavier.

\begin{figure}[t] 
   \centering
   \includegraphics[width=2.6in]{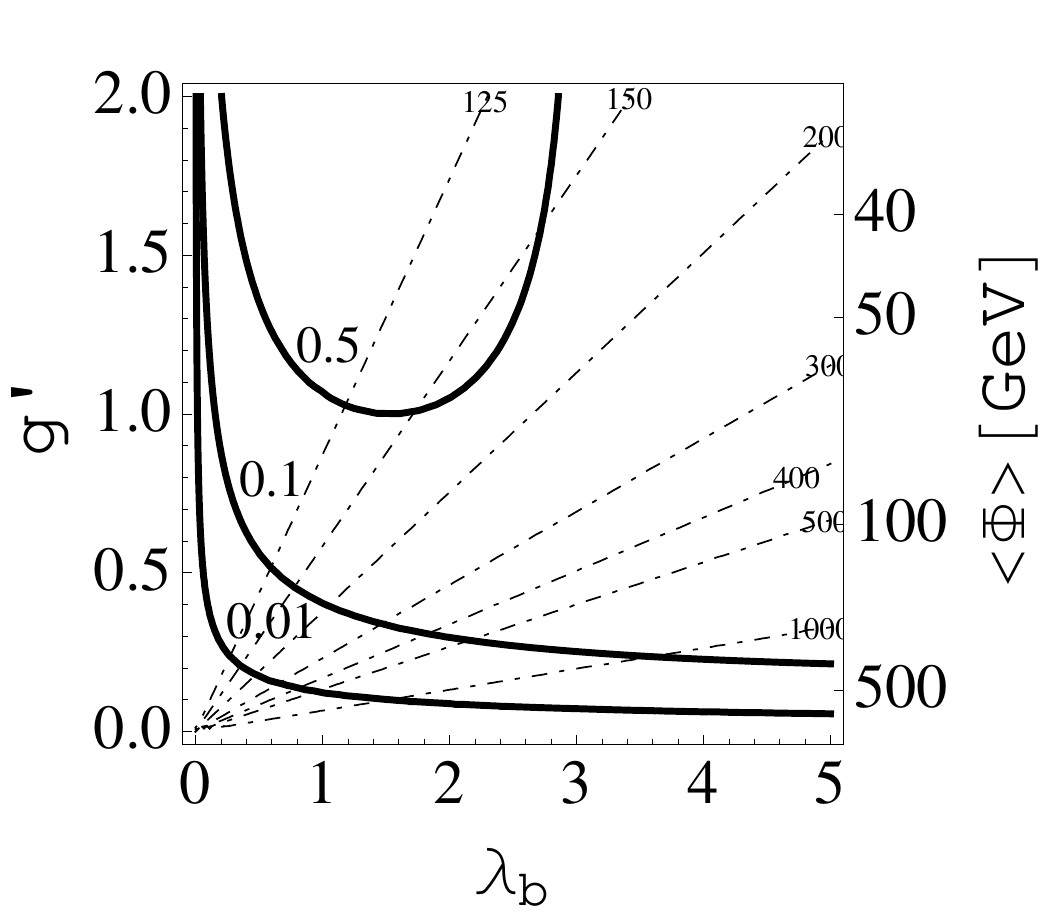}    
   \caption{Contours of  $g'^{bD}_R$ (solid lines) in the $g'$ - $\lambda_b$ plane for  $\mu_D = 200$ GeV. The rest as in Fig.~\ref{fig:g'lb}.}
   \label{fig:g'bB}
\end{figure}

For completeness, we also plot contours of constant $g'^{bD}_R$ in the $g'$ - $\lambda_b$ plane for  $\mu_D = 200$ GeV  in Fig.~\ref{fig:g'bB}.

 \subsection{Coupling of the electron to $Z'$}
 
The other required coupling besides $g'^b_R$ is  $g'^e_L$. This coupling can be generated in a very similar way, by adding 
a vector-like pair of fermions $L_L$ and $L_R$ charged under the $U(1)'$ where $L_L$ has the same quantum numbers under the standard model gauge symmetry as the lepton doublet $l_L$, see Table~\ref{tab:charges}. The $U(1)'$ charge assignment for heavy fermions and $\Phi$ is chosen so that heavy fermions fit into complete GUT multiplets, in this case $5$ and $\bar 5$ of $SU(5)$. The model is  anomaly free and its supersymmetric version preserves the gauge coupling unification of the standard model gauge couplings.

Our charge assignment results in the following renormalizable terms in the lagrangian:
\begin{equation}
{\cal L} \supset -  \bar l_{Li} Y^e_{ij} e_{Rj} H - \bar l_{Lk} \lambda^l_k L_{R} \Phi - \mu_L \bar L_L L_R + {\it h.c.}
\end{equation}
where the first term represents the usual standard model Yukawa couplings for charged leptons. The second term contains Yukawa interaction between lepton doublets  and the extra $L_R$ lepton. The last term is the mass term for the vector-like pair.  

The derivation of couplings in the charged lepton sector and the discussion of flavor violation closely follow the down quark sector. In the basis where standard model Yukawa  couplings are diagonal we choose $\lambda^l = (\lambda_e, 0, 0)$. This is the minimal case which generates $g'^e_L$ while the couplings of $\mu$ and $\tau$ to $Z'$ and flavor violating couplings are not generated. 

Due to opposite $Q'$ charges of $L_{R,L}$ and $D_{R,L}$, motivated by an SU(5) embedding, couplings  $g'^e_L$ and $g'^b_R$ have automatically opposite signs which is required by the best fit. The dependence of  
$g'^e_L$ on parameters of the model is identical to what we presented for $g'^b_R$, however the value of interest is much smaller. The value of $g'^e_L$ motivated by the best fit is about 1\% of the $g'^b_R$, see Table~\ref{tab:fit}.
This can be achieved for:
\begin{equation}
\frac{\lambda_b}{\mu_D} \simeq 10 \; \frac{\lambda_e}{\mu_L},
\end{equation}
which means that either the mixing coupling $\lambda_e$ is very small compared to $\lambda_b$ or the extra lepton $L$ is much heavier than the extra down quark $D$.
Since the mass of the electron is negligible, the generated  $g'^e_R$ and corrections to Z couplings are essentially zero.

 \subsection{Extensions of the model for other $Z'$ couplings and $Z$-$Z'$ mixing}

So far we have considered a  model that adds vector-like fields, with charges consistent with embedding into $5$ and $\bar 5$ of SU(5). Such a model  generates only $g'^b_R$ and  $g'^e_L$ couplings. The best fit with just these couplings is the fit II  in Table~\ref{tab:fit}.   Additional small couplings, $g'^{e}_R$ and $g'^{b}_L$, improve the quality of the fit somewhat.
Generating even sizable  $g'^{e}_R$ presents no challenge. However, $g'^{b}_L$ leads to a modification of both the 3rd row and 3rd column of the CKM matrix. More importantly these corrections are not suppressed by the mass of the $b$ quark and thus the generated $g'^{b}_L$ cannot be very large. However, the value of $g'^{b}_L$ suggested by the best fit to precision electroweak data is quite small,  see Fig.~\ref{fig:eR_and_bL} (Right), and even $g'^{b}_L = 0$ does not significantly change the fit. The   $g'^{b}_L$ is the least important coupling of the four.
 One can consider generating these additional small couplings by adding  vector-like fields with charges consistent with embedding into $10$ and $\bar {10}$ of SU(5).

Since the extra vector-like fermions couple to both $Z$ and $Z'$ their loops can generate $Z$-$Z'$ mixing. The contribution of a vector-like pair to the mixing can be however cancelled by adding a second vector-like pair with opposite $U(1)'$ charge. In addition, the mixing can be avoided when the $U(1)'$ is embedded into a non-abelian group.

\section{$Z'$ and $D$ at the Large Hadron Collider}

\begin{figure}[t] 
   \centering
   \includegraphics[width=2.in]{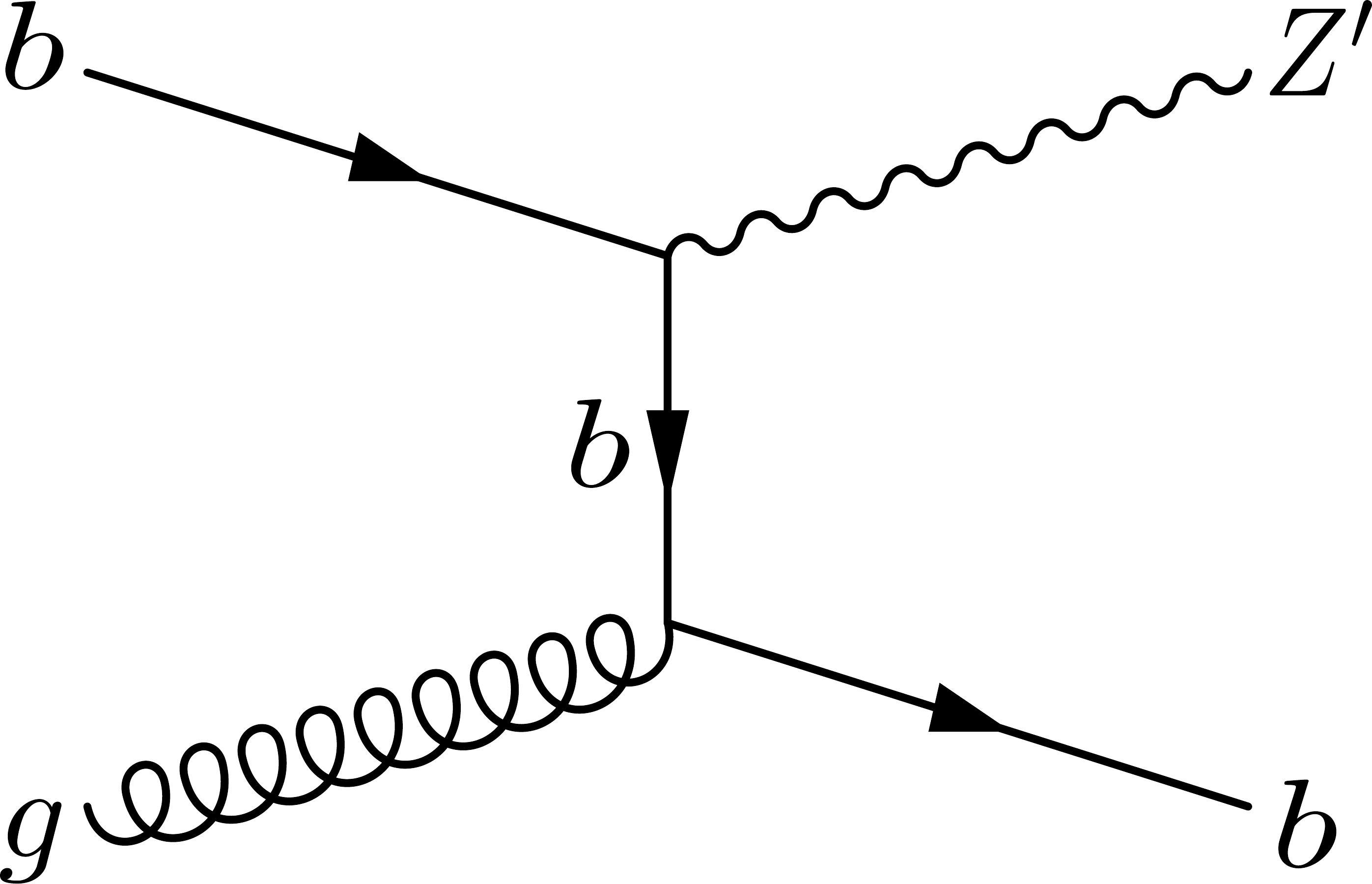}      
   \caption{Feynman diagram for $Z'$ production in association with the $b$ quark.}
   \label{fig:gb_to_Zpb}
\end{figure}

\begin{figure}[t] 
   \centering
   \includegraphics[width=2.3in]{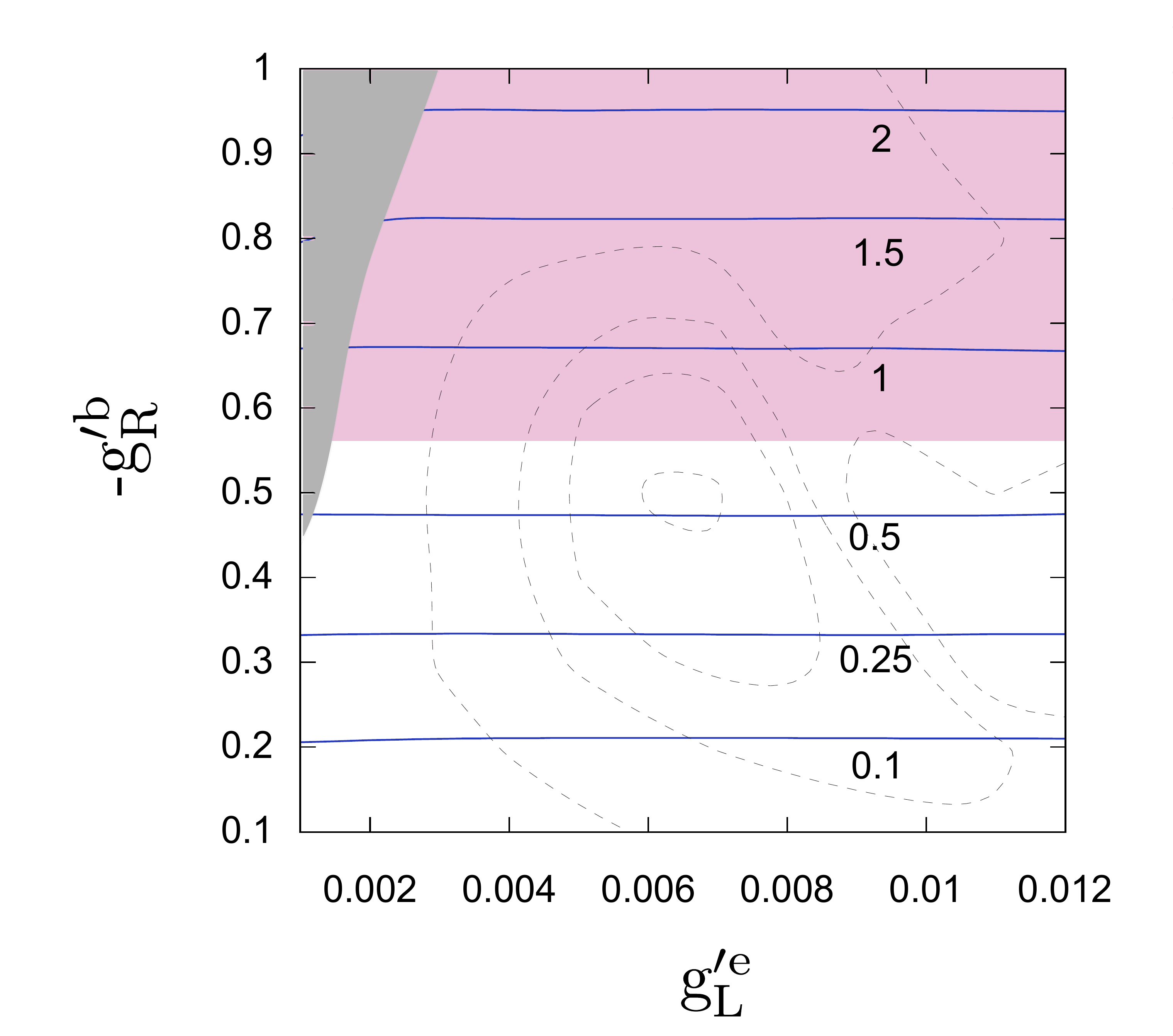}      
   \includegraphics[width=2.3in]{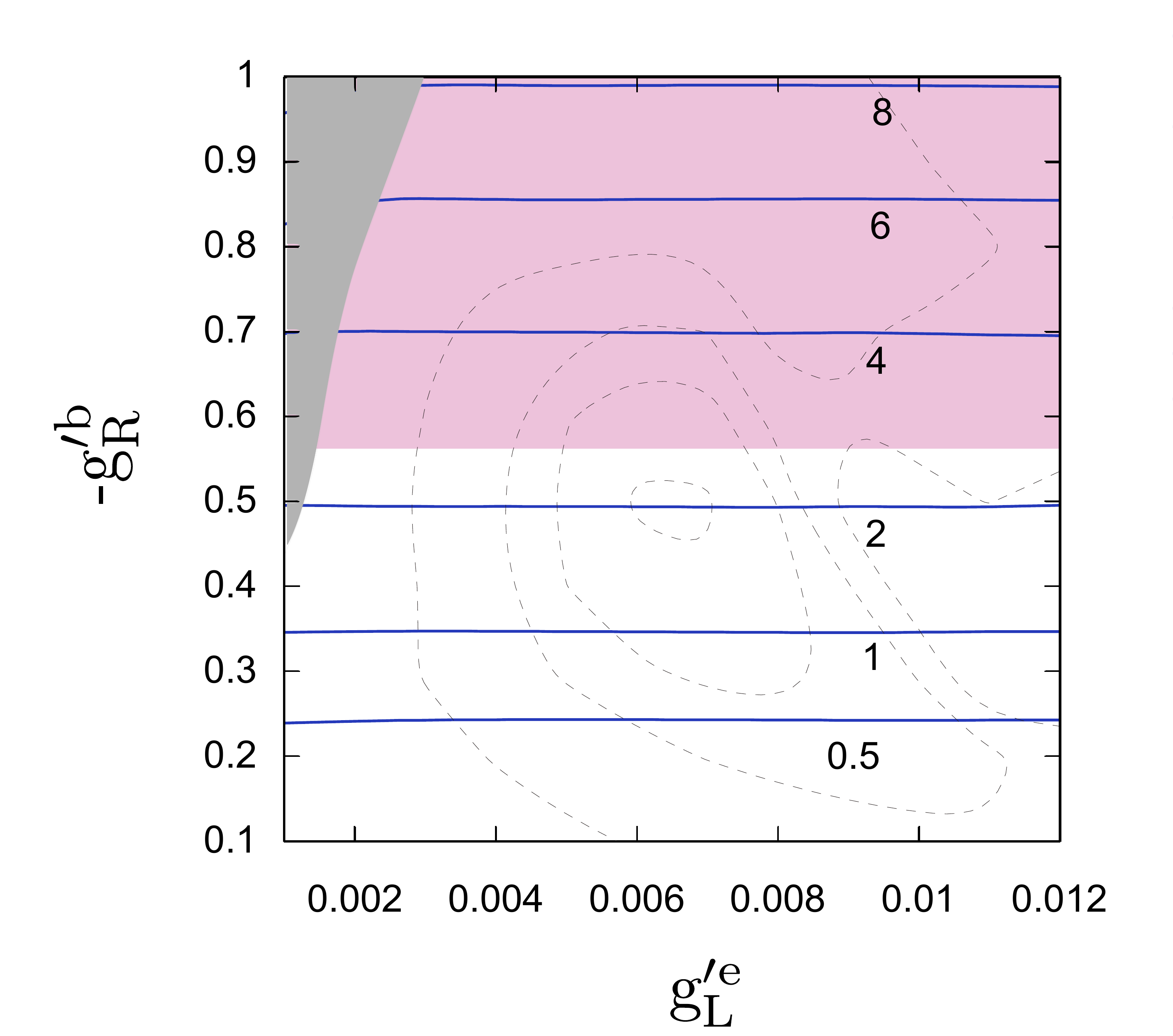}  
   \caption{$Z'b$ production cross section (nb) at the LHC with center-of-mass energy 7 TeV (Left) and 14 TeV (Right) with $\chi^2$ contours from Fig.~\ref{fig:chi2} overlayed. The shaded upper regions are excluded by the CDF search for the Higgs boson assuming $B(Z' \to b \bar b) = 100$\%.}
   \label{fig:Zpb}
\end{figure}

At hadron colliders the $Z'$ could be produced in association with $b$ quarks, See Fig.~\ref{fig:gb_to_Zpb}. 
The production cross sections of  $Z'b$ at the LHC are shown in Fig.~\ref{fig:Zpb} for center-of-mass energy of 7 TeV (Left) and 14 TeV (Right). 
The cross sections are calculated with MCFM \cite{Campbell:2010ff} at the leading order (LO).
We used CTEQ6.6 parton distribution functions (PDF).
The factorization and renormalization scales are set to $\mu_F = \mu_R = M_Z $.
For the final state b-jet, $p_T^b >15$ GeV, $|\eta| <2.5$ and $\Delta R < 0.7$ are chosen to match those used for the calculation of $Z b$ production which is  a background for Higgs searches~\cite{Campbell:2003dd}.
In the analysis no tagging efficiencies are assumed.

From Fig.~\ref{fig:Zpb} we see that the cross section is only governed by $g'^b_R$ since $g'^b_L$ is negligibly small.
In the region of the best fit the $Z'b$ cross sections are $\sim$0.5 nb at 7 TeV and $\sim$2 nb at 14 TeV. If other couplings besides $g'^{e,b}_{L,R}$ are absent the $Z'$ would decay to $b \bar b$ with branching ratio close to 100\%. The search for the $Z'$ is therefore very similar  to the search for the Higgs boson in 3b final state.

Recent limits on $\sigma(p \bar p \rightarrow \phi b) \times BR(\phi \rightarrow b\bar b)$ set by CDF with 2.6 fb$^{-1}$ of integrated luminosity~\cite{Aaltonen:2011nh} and D0 with 5.2 fb$^{-1}$~\cite{Abazov:2010ci} already constrain the allowed values of $g'^b_R$ .
We calculated the production cross sections of  $Z'b$ at the LO using MCFM with the center-of-mass energy of 1.96 TeV, $p_T^b >15$ GeV, $|\eta| <2$ and $\Delta R < 0.4$ that are used in the CDF search which currently gives  strongest limits.
Comparing it with the CDF limit $\sigma(p \bar p \rightarrow \phi b) \times BR(\phi \rightarrow b\bar b) \leq 26.4$ pb for $m_\phi = 90$ GeV, we find that $g'^b_R$ larger than 0.56 is excluded as shown in Fig.~\ref{fig:Zpb}. Note however that with possible couplings of $Z'$ to other quarks (or particles beyond the SM) the $BR(Z' \rightarrow b \bar b)$ can be highly reduced resulting in weaker limits.

At the LHC the $Z'b$ cross section is two orders of magnitude larger than at the Tevatron. 
So it is just a question of accumulating enough luminosity to see the signal of $Z'$. A search for the Higgs boson produced in association with the $b$ quark has not been performed yet at the LHC. 
Since predictions for the production cross sections depend on cuts used in an analysis,  let us make few comments. In the recent ATLAS measurement of the cross section for b-jets produced in association with a $Z$ boson decaying into two charged leptons, a b-jet was identified with $p_T^b >25$ GeV and $|y|<2.1$~\cite{Aad:2011jn}.
With these cuts on $p_T$ and $|y|$, the $Z'b$ production cross section is reduced to about half of those given in Fig.~\ref{fig:Zpb}.
Note also that MCFM is not interfaced to parton shower/hadronisation fragmentation package, and it does not include multiple parton interaction (MPI).
We expect about 10\% change in the cross sections given in  Fig.~\ref{fig:Zpb} once those corrections are taken into account \cite{Aad:2011jn}.
At the same time, the uncertainties stemming from the next-to-LO calculation, the scale dependences, PDF, and $\alpha_s$ are expected to be 20\%, 10\%, 3\%, and 2\%, respectively~\cite{Aad:2011jn}.

The model discussed in the previous section predicts the extra $D$ quark in a few hundred GeV range. Constraints from searches for the 4th generation do not apply, since the $D$ quark decays into $Z'b$ with $Z' \to b \bar b$. At the LHC the $D$ quark can be pair produced by QCD interactions leading to 6b final states. Since $Z' \to e^+ e^-$ is suppressed compared to $Z' \to b \bar b$ by $(g'^e_L/g'^b_R)^2 \simeq 10^{-4}$ the $e^+e^-4b$ final states are very rare.

\section{Conclusions and Outlook}

The $Z'$ near the $Z$-pole with couplings to the electron and the b-quark can resolve the puzzle in precision electroweak data by explaining the two largest deviations from SM predictions among $Z$-pole observables: $A_{FB}^b$ and  $A_e{(\rm LR-had.)}$. It nicely fits the energy dependence of $A_{FB}^b$ near the $Z$-pole and improves on  $\sigma_{\rm had}^0$ on the $Z$-pole and $R_b$ measured at energies above the $Z$-pole. 

We constructed a model that generates the minimal set of required couplings through mixing of standard model fermions with extra vector-like fermions charged under the $U(1)'$. It preserves standard model Yukawa couplings, it is anomaly free and can be embedded into grand unified theories. 
The model allows a choice of parameters that
 does not generate any flavor violating couplings of the $Z'$ to standard model fermions. Out of standard model couplings, it  only  negligibly modifies the left-handed bottom quark coupling to the $Z$ boson and the 3rd column of the CKM matrix. Modifications of standard model couplings in the charged lepton sector are also negligible. It predicts an additional down type quark, $D$, with mass in a few hundred GeV range, 
and an extra lepton doublet, $L$, possibly much heavier than the $D$ quark.

At the LHC the $Z'$ could be produced in association with b-quarks. The production cross sections of  $Z'b$ are large, in the region of the best fit as large as $\sim$0.5 nb for center-of-mass energy of 7 TeV and $\sim$2 nb for 14 TeV.
 If other couplings besides $g'^{e,b}_{L,R}$ are absent the $Z'$ would decay to $b \bar b$ with branching ratio close to 100\%. The search for the $Z'$ is therefore very similar  to the search for the Higgs boson in 3b final state. However,  with possible couplings of $Z'$ to other quarks (or particles beyond the SM) the $BR(Z' \rightarrow b \bar b)$ can be highly reduced which 
could make the search for $Z'$ difficult.  The optimal experiment to confirm or rule out the possibility of a $Z'$ near the $Z$-pole would be the future linear collider, especially the GigaZ option, which would allow more accurate exploration of the $Z$-peak.  

The extra $D$ quark can be pair produced at the LHC by QCD interactions. It dominantly decays into $Z'b$ leading to 6b final states. The $e^+e^-4b$ final states are highly suppressed.

Considering other flavor conserving couplings, or small flavor violating couplings,  expands the range of observables to which this $Z'$ could contribute. It would be interesting to see if it can simultaneously explain some other deviations from SM predictions. For example with additional couplings in the charged lepton sector the deviation in the muon $g-2$ can be explained~\cite{muon_g-2}. However, adding any additional couplings leads to many new constraints that have to be carefully examined.

\acknowledgements

We thank D. Bourilkov, J. de Blas, M. Gruenewald, H.D. Kim, R. Van Kooten, 
J. March-Russell, K. Moenig, and P. Langacker for useful discussions.   This work was supported in part 
by the award from Faculty Research Support Program at Indiana University.



\end{document}